\documentclass[aps,prx,showpacs,citeautoscript,superscriptaddress,longbibliography,notitlepage,twocolumn]{revtex4-1}

\usepackage{graphicx}  
\usepackage{epstopdf}
\usepackage{dcolumn}   
\usepackage{bm}        
\usepackage{amsmath,amssymb,amsfonts,bm,dsfont,units}
\usepackage{wrapfig}
\usepackage{array}
\usepackage[caption=false]{subfig}
\usepackage[bookmarks=false,linkcolor=blue,urlcolor=blue,colorlinks,citecolor=blue]{hyperref}
\usepackage{exscale,relsize}
\usepackage{verbatim}
\usepackage[usenames, dvipsnames]{color}
\newcommand{\ket}[1]{\left|#1\right>}
\newcommand{\bra}[1]{\left<#1\right|}

\begin{document}

\title{Anomalous exciton transport in response to a uniform, in-plane electric field }

\author{Swati Chaudhary}
\email{swatich@caltech.edu}
\affiliation{Department of Physics and Institute for Quantum Information and Matter, California Institute of Technology, Pasadena, CA 91125, USA}

\author{Christina Knapp}
\affiliation{Department of Physics and Institute for Quantum Information and Matter, California Institute of Technology, Pasadena, CA 91125, USA}
\affiliation{Walter Burke Institute for Theoretical Physics, California Institute of Technology, Pasadena, CA 91125, USA}

\author{Gil Refael}
\affiliation{Department of Physics and Institute for Quantum Information and Matter, California Institute of Technology, Pasadena, CA 91125, USA}
\affiliation{Walter Burke Institute for Theoretical Physics, California Institute of Technology, Pasadena, CA 91125, USA}

\date{\today}

\begin{abstract}
Excitons are neutral objects, that, naively, should have no response to a uniform, electric field.
Could the Berry curvature of the underlying electronic bands alter this conclusion? 
In this work, we show that Berry curvature can indeed lead to anomalous transport for excitons in 2D materials subject to a uniform, in-plane electric field.
By considering the constituent electron and hole dynamics, we demonstrate that there exists a regime for which the corresponding anomalous velocities are in the same direction.
We establish the resulting center of mass motion of the exciton through both a semiclassical and fully quantum mechanical analysis, and elucidate the critical role of Bloch oscillations in achieving this effect.
We identify transition metal dichalcogenide heterobilayers as candidate materials to observe the effect.
\end{abstract}

\maketitle

\section{Introduction}

Berry curvature of electronic bands plays an important role in the transport phenomena and optical responses of a system~\cite{Xiao10}.
Among the myriad consequences of a finite Berry curvature is the anomalous velocity, in which an electron experiencing a force perpendicular to the Berry curvature of the band acquires a contribution to the velocity perpendicular to both.
The anomalous velocity can be well understood from a single particle and semi-classical treatment, and leads to a variety of interesting features including the quantum anomalous Hall effect. 
In this work, we consider the role played by the anomalous velocity for exciton transport.

Excitons have attracted renewed interest for their dominant role in the optical response of van der Waals materials~\cite{Fogler14,Rivera16,Chen16,Nagler17,Wang18,Donck18, Kunstmann18,Merkl19,Miller17}.
An exciton is a neutral boson consisting of an electron-hole pair bound by Coulomb interactions.
In van der Waals materials, such as transition metal dichalcogenides (TMDs), excitons exhibit a variety of interesting behaviors intimately tied to Berry curvature, including valley selective optical response~\cite{Yao08b,Cao12,Mak12, Mak14,Jin18}, topological bands in the presence of a moir\'e potential~\cite{Sie15,Wu17, Wu18, Kwan20}, and non-hydrogenic spectra~\cite{Srivastava15,Zhou15,Alloca18,Hichri19}. 
In particular, anomalous exciton transport in response to electric and magnetic fields has garnered significant interest~\cite{Kovalev19,Jauregui17,Kozin06,Onga17}. 
Typically, such transport requires a net force acting on the exciton center of mass, e.g. by utilizing the exciton dipole moment. 
In contrast, here we consider excitons confined to a two dimensional system in the presence of a uniform in-plane electric field.
Given the absence of a net force on the exciton center of mass, anomalous transport can only arise by considering the internal structure of the exciton.

Heuristically, one might anticipate that in response to a uniform, in-plane electric field the electron and hole composing the exciton would initially move apart until they reach an equilibrium point at which the force from the electric field is balanced by the Coulomb interaction.
If the electronic bands have some out-of-plane Berry curvature component, both constituent particles will move with an anomalous velocity while they experience a net force~\cite{Xiao10}.
In the case of intervalley excitons, the electron and hole bands can experience the same Berry curvature.
As a result, the anomalous velocity will point in the same direction for the electron and hole, thereby resulting in exciton center of mass motion.
However, this anomalous motion will only happen for the short period of time that it takes the electron and hole to reach their equilibrium separation, after which the exciton will once again remain stationary.

In this work, we show that when the electron and hole undergo Bloch oscillations, the anomalous velocity persists over an extended period of time resulting in a measurable anomalous exciton drift
While the effect is predicted to be stronger for electronic bands, it does not require them.
Essentially, Bloch oscillations bound the relative separation of the electron and hole by the bandwidth so that the electron and hole can never reach their equilibrium separation.
The exciton center of mass moves as a result of the sustained anomalous velocity, resulting in anomalous transport in response to a uniform in-plane electric field.
A semiclassical analysis predicts Bloch oscillations when the electric field is sufficiently large compared to the Coulomb interaction.
Surprisingly, we find that Bloch oscillations can also occur when the interaction strength is large compared to the electric field.
This additional region of parameter space supporting anomalous exciton transport only manifests when evolving the exciton quantum mechanically.

The remainder of this work is organized as follows.
In Section~\ref{sec:theory}, we identify the necessary ingredients for anomalous exciton transport.
We first consider a semiclassical analysis and derive a lower bound on the electric field for the electron and hole to experience Bloch oscillations. 
We then motivate why the small-field, strong-interaction limit also supports Bloch oscillations, with insight from a simple 1D toy model.
In Section~\ref{sec:numerics}, we present numerically simulated phase diagrams of the anomalous exciton drift when the underlying electronic bands are both topological and trivial.
We plot the semiclassical equations of motion for both harmonic and Coulombic potentials. 
We further simulate exciton dynamics quantum mechanically for a toy model of the electron and hole, again with both harmonic and Coulombic potentials.
Section~\ref{sec:candidate-systems} identifies additional complications beyond the models considered in the previous sections and argues that TMD heterobilayers are an attractive candidate system to observe anomalous exciton transport.
Finally, in Section~\ref{sec:discussion} we discuss the relation to previous works and identify future directions.
Details of the analytical and numerical analyses are relegated to the appendices.

\begin{figure}
    \centering
    \includegraphics[scale=0.1]{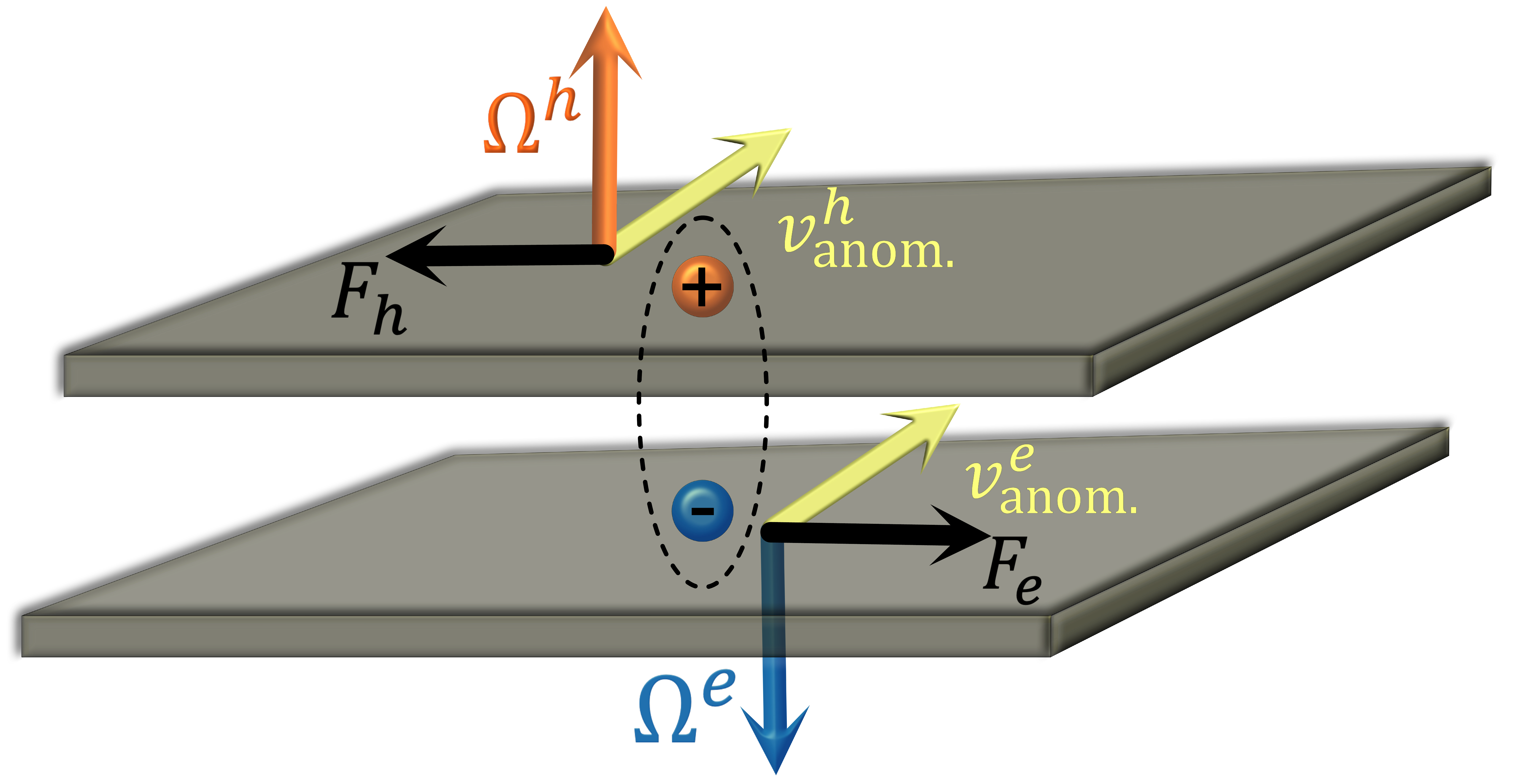}
    \caption{Schematic of an interlayer exciton with electron and hole bound to opposite layers.  When the valence and conduction bands have the same Berry curvature, the anomalous velocity of the electron and hole $v^{e/h}_\text{anom}$ points in the same direction, resulting in a net transverse drift of the exciton.}
    \label{fig:schematic}
\end{figure}

\section{Anomalous exciton drift: theory}\label{sec:theory}

In this section, we introduce the toy models used to study anomalous exciton transport.
We begin with a semiclassical analysis highlighting the critical role of Bloch oscillations.
We then motivate how a quantum mechanical treatment indicates an additional region supporting the effect that is not predicted by semiclassics.

\subsection{Semiclassical exciton dynamics}\label{sec:semiclassics}

We model the exciton as an electron-hole pair subject to an interaction potential $U(\mathbf{r}_e-\mathbf{r}_h)$ with the electron in the conduction band and the hole in the valence band.
We assume both bands have finite Berry curvature $\Omega^{c/v}_{e/h}$.
In the presence of a uniform in-plane electric field, the semiclassical equations of motion are
\begin{align}
\dot{\mathbf{k}}_{e/h} &= -{\bf \nabla}_{\mathbf{r}_{e/h}} U(\mathbf{r}_e-\mathbf{r}_h) \mp  e \mathbf{E} \label{eq:semi-k}
\\ \dot{\mathbf{r}}_{e/h} &= \partial_{\mathbf{k}_{e/h}} \varepsilon_{e/h}(\mathbf{k}_{e/h}) - \dot{\mathbf{k}}_{e/h} \times \mathbf{\Omega}^{c/v}_{e/h} \label{eq:semi-r}
\end{align}
where we have set $\hbar=1$ and $\varepsilon_{e/h}(\mathbf{k}_{e/h})$ denotes the electron/hole band dispersion. 
In Eq.~\eqref{eq:semi-r}, the first term corresponds to the group velocity resulting from the band dispersion, while the second term is the anomalous velocity resulting from finite Berry curvature.

The force experienced by the electron and hole are precisely opposite for uniform $\mathbf{E}$.
As such, the force on the exciton center of mass is strictly vanishing,
\begin{align}
    \dot{\mathbf{K}}&=  \dot{\mathbf{k}}_e + \dot{\mathbf{k}}_h = 0.
\end{align}
The relative and center of mass (COM) position coordinates evolve according to
\begin{align} \label{eq:sm-eom-r}
 \dot{\mathbf{r}} &=  \partial_{\mathbf{k}_e} \varepsilon_e (\mathbf{k}_e)-\partial_{\mathbf{k}_h} \varepsilon_h (\mathbf{k}_h)- \dot{\mathbf{k}}_e \times\left(\mathbf{\Omega}^c_e (\mathbf{k}_e)+\mathbf{\Omega}^v_h(\mathbf{k}_h)\right)
\\ \dot{\mathbf{R}} &= \frac{ \partial_{\mathbf{k}_e} \varepsilon_e (\mathbf{k}_e)+\partial_{\mathbf{k}_h} \varepsilon_h (\mathbf{k}_h)}{2}
 - \dot{\mathbf{k}}_e \times \frac{\mathbf{\Omega}^c_e(\mathbf{k}_e)-\mathbf{\Omega}^v_h(\mathbf{k}_h)}{2}\label{eq:sm-eom-R}
\end{align}
when $m_e=m_h$.
Note that for a given band $\alpha$, $\mathbf{\Omega}^\alpha_h(\mathbf{k}_h)=-\mathbf{\Omega}^\alpha_e(-\mathbf{k}_h)$, see Appendix~\ref{app:derivation} for a derivation.

In a particle-hole symmetric two-band model, ${\varepsilon_e(\mathbf{k}_e)=\varepsilon_h(-\mathbf{k}_h)}$ 
and ${\mathbf{\Omega}_e^c(\mathbf{k}) = \mathbf{\Omega}_h^v(\mathbf{k})}$.
Therefore, a direct momentum exciton $\mathbf{k}_e=-\mathbf{k}_h$ has no center of mass motion, $\dot{\mathbf{R}}=0$.
In this case, the Berry curvature can only affect the relative motion of the electron and hole.
These internal dynamics can affect the exciton spectra~\cite{Srivastava15,Zhou15}, but do not result in anomalous transport.

In contrast, any deviation from the two band, direct-momentum, particle-hole symmetric system can result in Berry curvature effects on COM motion.
Motivated by intervalley excitons in TMD bilayers, we consider a direct momentum exciton $\mathbf{k}_e=-\mathbf{k}_h$ with opposite Berry curvatures for the conduction and valence bands~\footnote{We take momentum measured from the Dirac point to consider intervalley excitons and $\mathbf{k}_e=-\mathbf{k}_h$ simultaneously.}.
The corresponding relative and COM equations are 
\begin{align}
\dot{\mathbf{r}} &= 2\partial_{\mathbf{k}_e} \varepsilon_e (\mathbf{k}_e), &\dot{\mathbf{R}} &= - 2\dot{\mathbf{k}}_e \times\mathbf{\Omega}^c_e(\mathbf{k}_e).
\end{align}
In the absence of interactions, the relative and center of mass motions decouple, and an electric field $\mathbf{E}$ results in a net transverse drift of the exciton if the line integral of Berry curvature along the direction of $\mathbf{E}$ is non-zero.

Interactions complicate the story by coupling the relative and center of mass motion.
The relative strength of interactions, bandwidth, and electric field result in two limiting regimes:
\begin{enumerate}
	\item {\bf Harmonic oscillator regime}: 
	The restoring force is able to overcome the applied electric field and the relative momentum $\mathbf{k}$ of the exciton is not able to reach the Brillouin zone boundaries. 
	As a result, $\dot{\mathbf{k}}$ changes sign and the exciton oscillates perpendicular to the direction of $\mathbf{E}$.
	\item {\bf Bloch oscillation regime}: 
	The electric field is sufficiently large that $\dot{\mathbf{k}}$ has the same sign as $\mathbf{E}$ at all times.
	The relative momentum of the exciton crosses the Brillouin zone momentum-space boundary.
	The cooperative anomalous velocity results in a non-vanishing transverse drift in center of mass position whenever the line-integral of Berry curvature along the direction of $\mathbf{E}$ is non-zero.  
\end{enumerate}
The restoring force is provided by the attraction between the electron and hole; as such, it depends on both the strength of interactions (e.g. dielectric constant) and the displacement from the equilibrium position. 
Below, we model the electron-hole interaction with a simple harmonic potential, before considering the more realistic Coulombic potential.

\subsubsection{Harmonic Potential}

When the attraction between the electron and hole is modeled as a harmonic potential
\begin{align}\label{eq:harmonic-U}
    U(\mathbf{r})=-V_0+\frac{1}{2}\kappa r^2,
\end{align}
 the non-vanishing equations of motion in the presence of a uniform electric field $\mathbf{E}=E\mathbf{\hat{x}}$ are 
\begin{align}
\dot{\mathbf{r}}(t) &=2Ja\sin\left( k_x(t)a\right)\,\hat{\mathbf{x}}+2Ja\sin\left( k_ya\right)\,\hat{\mathbf{y}} \label{eq:8} \\
\dot{\mathbf{k}}(t) &=- \left(\kappa x(t)+e E\right)\,\hat{\mathbf{x}}-\kappa y(t)\,\hat{\mathbf{y}} \label{eq:9} \\
\dot{\mathbf{R}}(t) &=2\dot{k}_x\Omega^c_e(\mathbf{k}(t))\, \mathbf{\hat{y}}-2\dot{k}_y\Omega^c_e(\mathbf{k}(t))\mathbf{\hat{x}} \label{eq:10}
\end{align}
where $\mathbf{r}=\mathbf{r}_e-\mathbf{r}_h$ and $\mathbf{k}=(\mathbf{k}_e-\mathbf{k}_h)/2$.
If there is no interaction, we expect Bloch oscillations in relative space with amplitude and period 
\begin{align}\label{eq:x-Bloch}
    x_\text{Bloch}&=\frac{2J}{eE}, & \tau_\text{Bloch}&=\frac{2\pi}{eEa}.
\end{align}
The exciton COM experiences a net transverse drift provided $\oint\Omega^c_e(\mathbf{k})dk_x\neq 0.$

Interactions reduce the magnitude of $\mathbf{\dot{k}}(t)$, and thus slow down the anomalous velocity.
If at any point $x(t)$ exceeds the equilibrium position $x_\text{eq}=eE/\kappa$, the electron/hole does not reach the Brillouin zone boundary and $\mathbf{\dot{k}}(t)$ changes sign.
In this case, both relative and COM motion oscillate, corresponding to the harmonic oscillator regime.

The value of $E$ for which $x_\text{eq}>x_\text{Bloch}$ sets a lower bound on $E$ to achieve Bloch oscillations; in the absence of interactions, this bound is given by
\begin{align}
    eE&>\sqrt{2J\kappa}.
\end{align}
Interactions modify the above, but do not change the fact that semiclassics only predicts anomalous exciton transport when the electric field is sufficiently large compared to $J$ and $\kappa$.

Thus far, we have focused on the simple limit ${\mathbf{\Omega}^c_e(\mathbf{k})=\mathbf{\Omega}^v_e(\mathbf{k})}$ for which the relative and COM motion decouple, with Berry curvature affecting only the latter.
In the opposite limit $\mathbf{\Omega}^c_e(\mathbf{k})=-\mathbf{\Omega}^v_e(\mathbf{k})$,
relative and COM motion again decouple, with Berry curvature affecting only the former.
More generally, both the relative and COM motion will be coupled by the Berry curvature terms.
We discuss this intermediate case in Appendix~\ref{app:intermediate}.

\subsubsection{Coulombic Potential}

More realistically, we expect the electron and hole to be bound by Coulomb interactions. 
For interlayer excitons confined to layers separated by a distance $D$,
\begin{align}
  U(\mathbf{r})=-\frac{k e^2}{\epsilon}\frac{1}{\sqrt{D^2+r^2}}= - \frac{\kappa D^2}{\sqrt{1+r^2/D^2}},  
\end{align}
where $\kappa\equiv ke^2/(\epsilon D^3).$
When $r\ll D$, $U(\mathbf{r})$ is well approximated by Eq.~\eqref{eq:harmonic-U} with $V_0=\kappa D^2$. 

The corresponding equations of motion for relative and COM position are again given by Eqs.~\eqref{eq:8} and \eqref{eq:10}, but $U(\mathbf{r})$ modifies Eq.~\eqref{eq:9} to
\begin{align}
\dot{\mathbf{k}}(t) &=-\nabla_{\mathbf{r}} U(\mathbf{r})-eE\,\hat{\mathbf{x}}.
\end{align}
We again expect a net transverse drift in exciton COM motion for sufficiently large $E$.
When $\langle x\rangle \ll D$, the transition between Bloch and harmonic oscillator regimes should agree with the bound derived for a harmonic potential.
When $\langle x \rangle \sim D$, the restoring force is weaker than for the harmonic potential case.
We expect this to result in the anomalous drift persisting for a larger region of $E$ versus $\kappa$ space.

Our semiclassical analysis considers separate wavepackets for the electron and hole.
Alternatively, the single particle semiclassical formalism can be extended for an exciton, as was done recently by Ref.~\onlinecite{Cao20}. 
We compare these approaches in Appendix~\ref{app:scdynamics}.

\subsection{Small field limit: intuition from 1D toy model}\label{sec:quantum}

The semiclassical analysis predicts anomalous exciton transport only when the electric field $E$ is sufficiently large compared to the interaction strength $\kappa$.
Our numerics, however, indicate that there is also anomalous drift in the small field regime. 
We can gain insight into this regime by considering a 1D toy model
\begin{align}\label{eq:H-1D}
H_\text{1D}&=-\sum_n  \left[ J\left(\ket{n}\bra{n+1}+\text{h.c}\right)+\frac{1}{2}\kappa a^2 \,\hat{n}^2\ket{n}\bra{n}\right] \notag 
\\ &\quad + e Ea \sum_n \hat{n}\ket{n}\bra{n},
\end{align}
where $\ket{n}$ corresponds to the position eigenstate on the $n$th lattice site and $\hat{n}\ket{n}=n\ket{n}$.
The position $n$ represents the relative coordinate of the electron and hole in the exciton discussion.

In the previous section, we argued that for appropriate Berry curvature profiles, the exciton experiences an anomalous drift when the electron and hole cannot reach their equilibrium separation.
The analogous consideration for the 1D toy model in Eq.~\eqref{eq:H-1D} is to consider when the position expectation value $\langle a \hat{n}\rangle$ is less than the equilibrium separation $x_\text{eq} \sim eE/\kappa.$
If we begin in the ground state of $H_\text{1D}$ for $E=0$ and evolve for finite $E$, we find two regimes.
When $\kappa a^2/J\lesssim 1$, the ground state resembles a wavepacket in both position and momentum space, resulting in good agreement with the semiclassical dynamics.  
In contrast, when $\kappa a^2/J\gg 1$, the ground state wavefunction $\psi_0(x)$ is confined to a single site and is therefore spread over the full Brillouin zone.
The wavefunction experiences an averaged group velocity, resulting in a much smaller restoring force compared to the semiclassical regime.
The position expectation value $\langle a \hat{n}\rangle$ oscillates with amplitude $\sim \left(J^2/(\kappa^2 a^4)\right) x_\text{eq} \ll x_\text{eq}$, see Appendix~\ref{app:1D}, allowing Bloch oscillations even in the small $E$ field limit.
Extrapolating to exciton dynamics, we should therefore expect anomalous drift in both the semiclassical Bloch oscillation regime {\it and} in the small field-large interaction limit.
We emphasize that the latter required taking into account the finite spread of the wavefunction in position and momentum space, and thus only emerges in a quantum mechanical treatment of the dynamics.

\section{Anomalous Exciton Drift: Numerics}\label{sec:numerics}

We now numerically simulate a toy model of an exciton whose electron and hole occupy bands with the same Berry curvature profile.  
We consider both a semiclassical and a quantum mechanical model with similar band dispersion and Berry curvature profiles.
For each case, we consider both harmonic and Coulombic potentials.

\subsection{Semiclassical numerics}

\begin{figure}
	\includegraphics[scale=0.5]{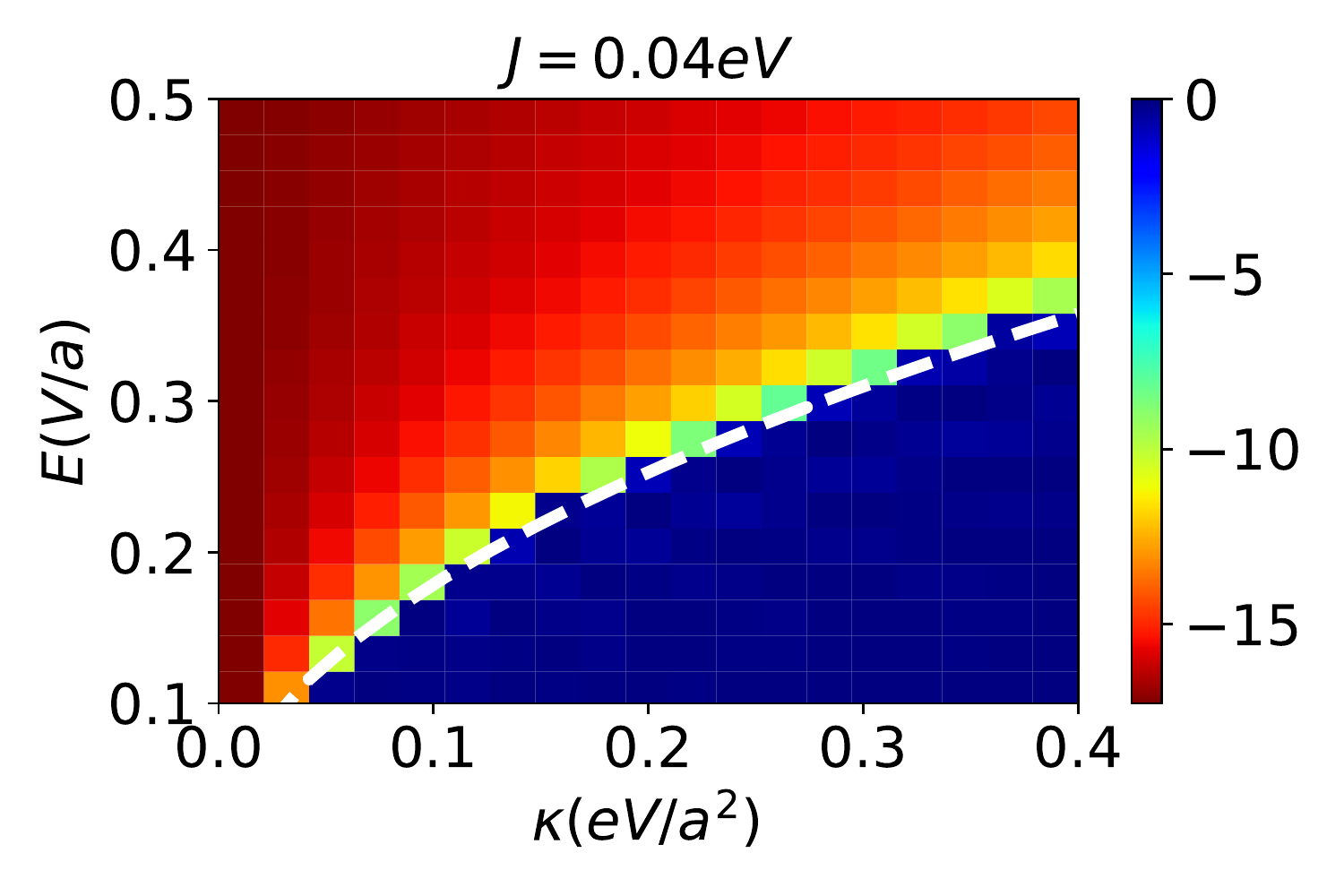}
	\includegraphics[scale=0.5]{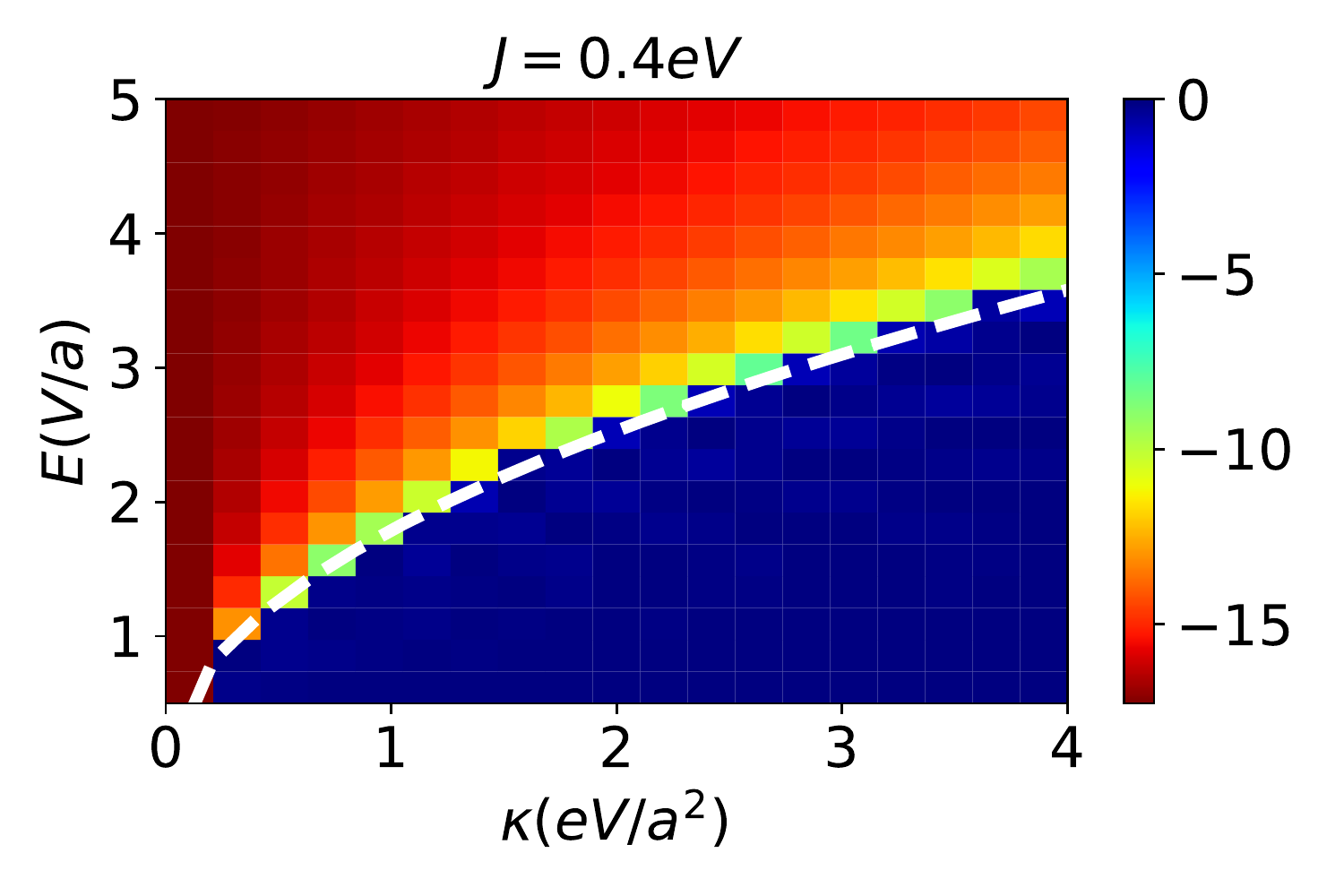}
	\caption{{\it Semiclassical dynamics for harmonic potential.} Average $Y$ per Bloch cycle in units of the lattice constant $a$ plotted against $E$ and $\kappa$ for $J=0.04$\,eV (top) and $J=0.4$\,eV (bottom). 
	The white dashed curve shows $2\sqrt{2J\kappa}$ separating the harmonic oscillator regime (dark blue) from the Bloch oscillator regime (red).
	Note that the $x$-axis is in units of eV/$a^2$ and $y$-axis is in units of V/$a$, where $a$ is the lattice constant.  For $E=0.25$\,V/a, $a=8$ nm, a transverse displacement of $5a$ indicates an anomalous velocity, $v_\text{anom.}^y\approx 3\times10^6$ m/s.
    }
	\label{fig:sc}
\end{figure}

We simulate the center of mass motion according to the semiclassical equation of motion Eq.~\eqref{eq:sm-eom-R} with a simple cosine dispersion and with the Berry curvature profile set by the BHZ Hamiltonian~\cite{Bernevig06}.
More explicitly, we take the electron and hole to evolve according to the upper band of the band-flattened Hamiltonian
\begin{align}\label{eq:H-FB}
    H_\text{BHZ}^\text{FB}(\mathbf{k})  &= \varepsilon(\mathbf{k}) \frac{ H_\text{BHZ}(\mathbf{k})}{\mathcal{E}_\text{BHZ}(\mathbf{k})}.
\end{align}
In the above, $H_\text{BHZ}$ and $\mathcal{E}_\text{BHZ}$ are the BHZ Hamiltonian and energy spectrum, 
\begin{align} \label{eq:H-BHZ}
    H_\text{BHZ}(\mathbf{k})&=\sum_{j\in \{x,y,z\}} c_j(\mathbf{k}) \sigma_j 
    \\ \mathcal{E}_\text{BHZ}(\mathbf{k}) &= \sqrt{ c_x(\mathbf{k})^2 + c_y(\mathbf{k})^2 + c_z(\mathbf{k})^2},
\end{align}
for $c_z = m_0 -b[\cos(k_x a) + \cos(k_y a)]$, $c_{x/y} = v_{x/y} \sin(k_{x/y} a)$.
We take the dispersion $\varepsilon$ to be
\begin{align}
    \varepsilon (\mathbf{k})=-J\left(\cos \left(k_xa\right)+\cos \left(k_ya\right) \right).
\end{align}

Figure~\ref{fig:sc} plots the average transverse COM motion per Bloch cycle in units of the lattice constant $a$.
We see that in the presence of harmonic interactions, the semiclassical Bloch oscillation regime is bounded by 
\begin{equation}\label{eq:sm-limit}
eE>2\sqrt{2J\kappa}.
\end{equation} 
Above this bound, the exciton experiences an anomalous drift; below it, the exciton's center of mass displacement averages to zero.
The top and bottom panels correspond to different values of the bandwidth $J$; as expected, the phase diagram is unchanged by scaling $J$, $\kappa$, and $E$ by the same factor.
We take parameters $m_0=1.4$\,eV, $b=1$\,eV, $v_x=\pm v_y^{e/h}=0.9$\,eV, and set $e=1$, corresponding to topological electronic bands.
With these parameters and $ E=0.25$\,V/$a$, $a=8$\,nm, a transverse displacement of $5a$ indicates an anomalous velocity $v_\text{anom}^y\approx 3\times 10^6$\,m/s.

\begin{figure}
	\includegraphics[scale=0.5]{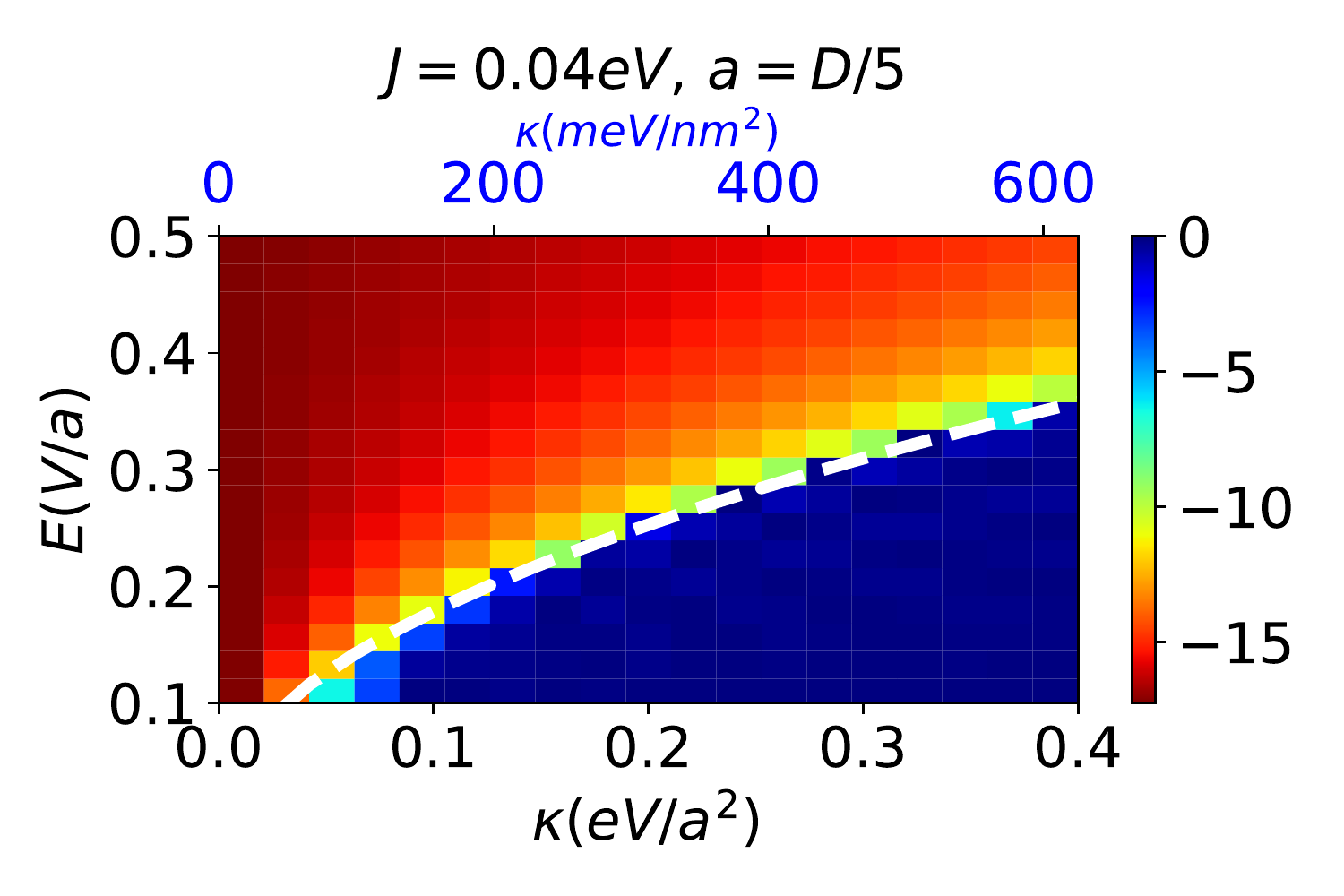}
	\includegraphics[scale=0.5]{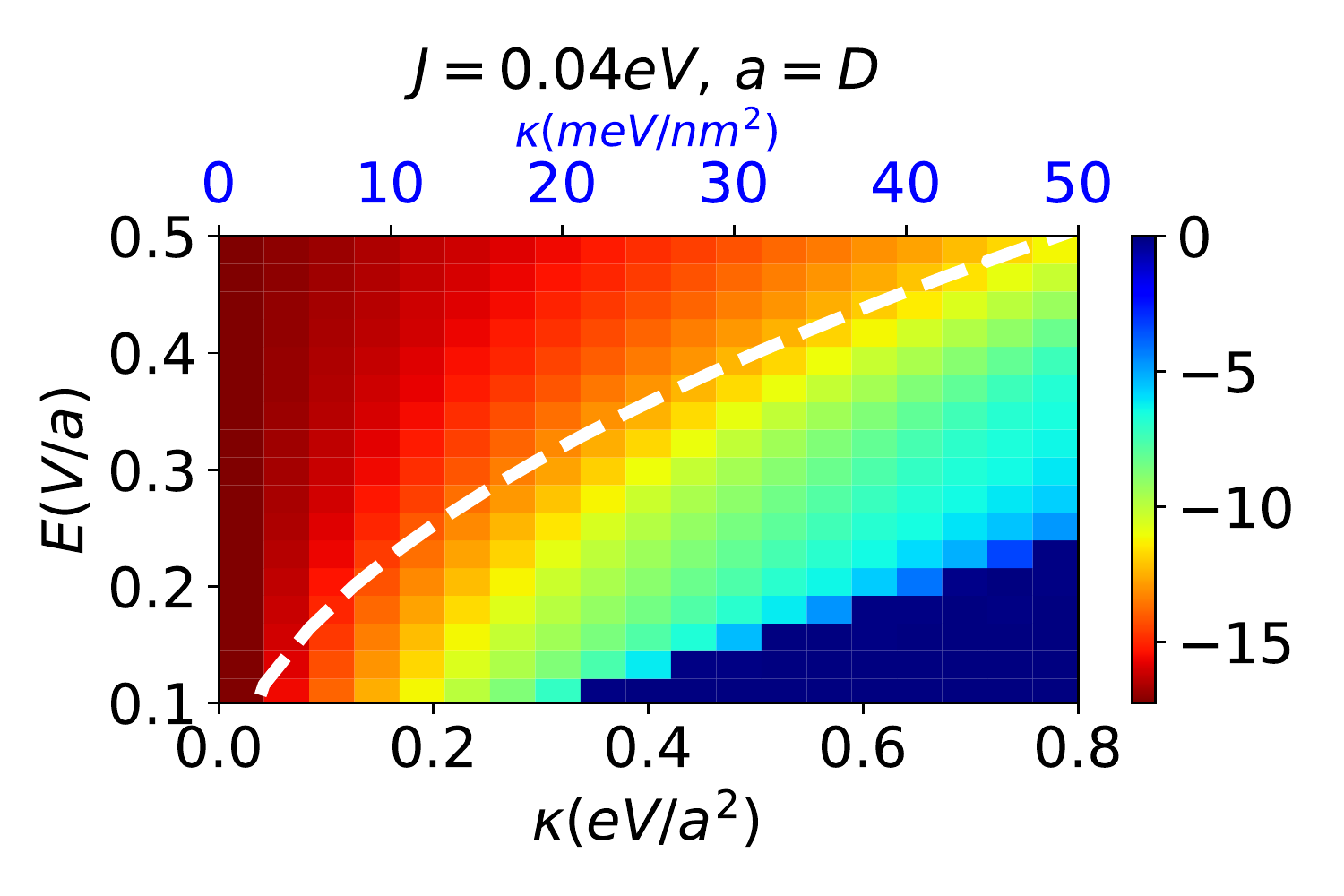}
	\caption{{\it Semiclassical dynamics for Coulombic potential.} 
	Same phase diagram as in Fig.~\ref{fig:sc} for Coulombic interaction with $J=0.04$\,eV.
	For $a\ll D$ (top) the phase diagram is similar to that in Fig.~\ref{fig:sc}.
	For $a=D$ (bottom), there is an increased Bloch oscillation regime compared to the harmonic potential case. 
	The top $x$-axis in blue indicates the absolute scale of $\kappa$ in meV/nm$^2$.
	The Berry curvature profile is again similar to that of Eq.~\eqref{eq:H-BHZ} with the same parameters as in Fig.~\ref{fig:sc}.}
	\label{fig:scCoulombic}
\end{figure}

We plot the same phase diagram for Coulombic potential in Fig.~\ref{fig:scCoulombic}.
The white dashed curve again plots the bound in Eq.~\eqref{eq:sm-limit}.
We see that for large $D$ (top), the plot agrees with the phase diagram for the harmonic potential.
As anticipated, for small $D$ the Bloch oscillation regime extends beyond this bound.
We take the same parameters as for Fig.~\ref{fig:sc}.

\subsection{Exact dynamics simulation}

\begin{figure}
	\includegraphics[scale=0.5]{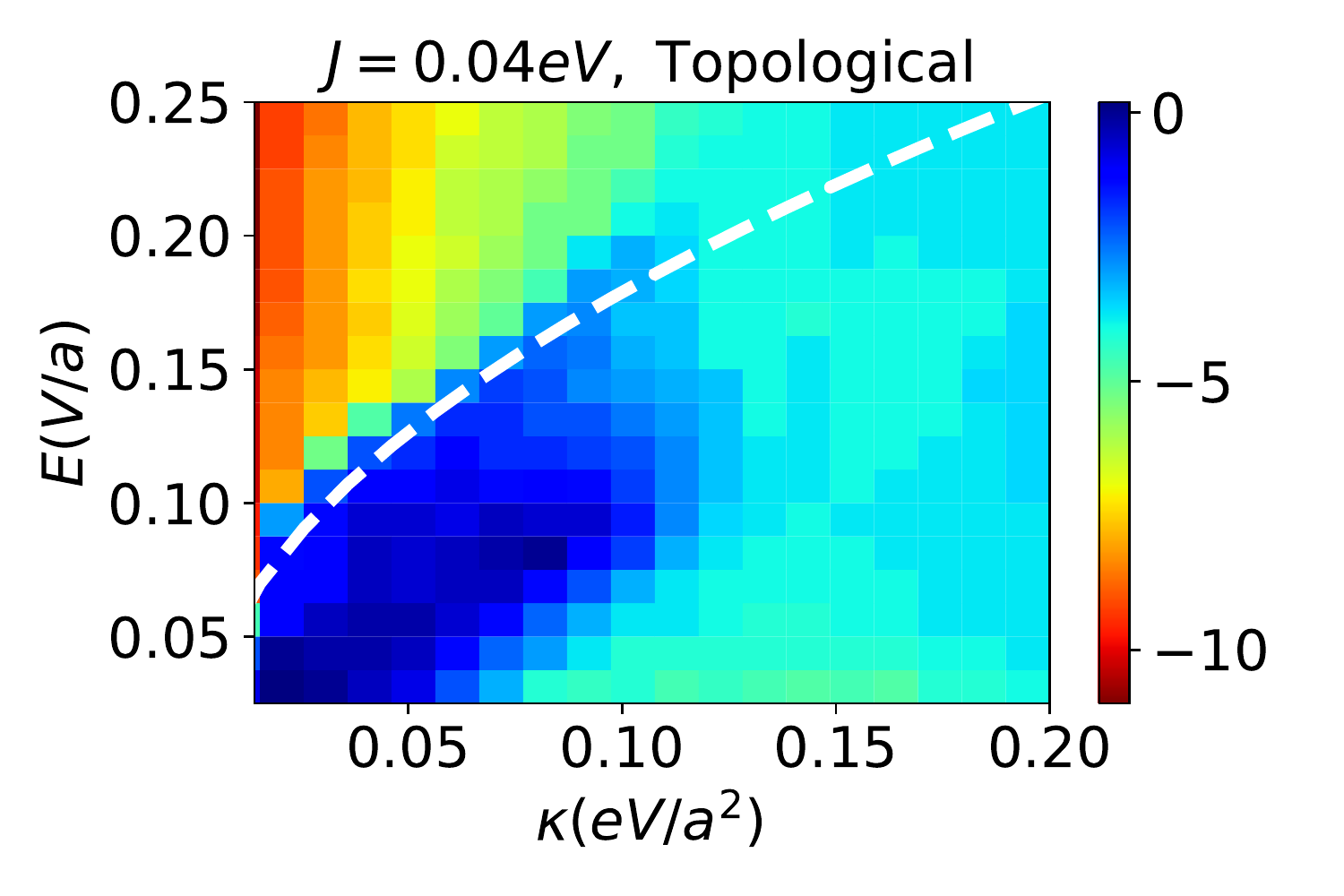}
	\includegraphics[scale=0.5]{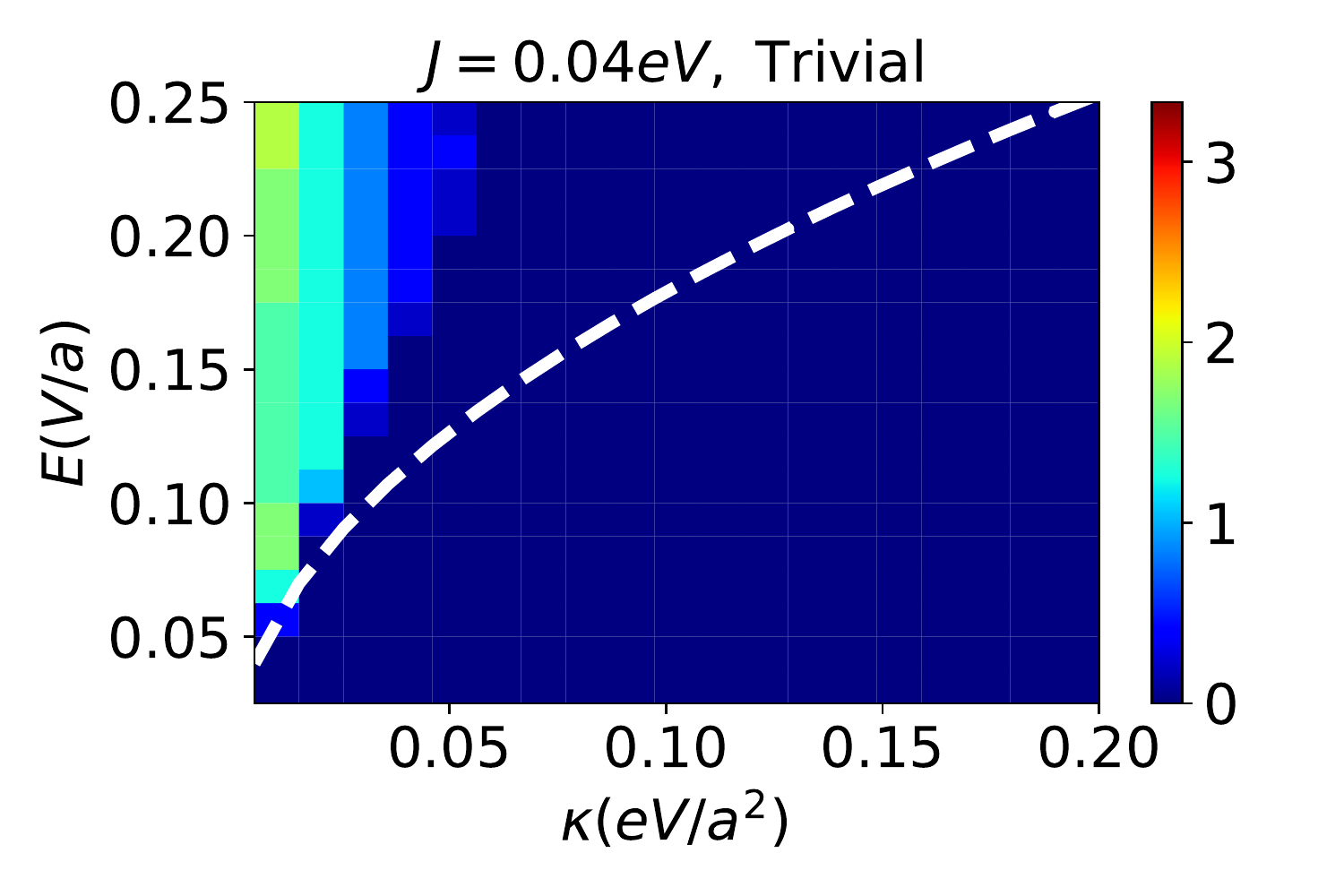}
\caption{{\it Exact dynamics for harmonic potential}.
    Same phase diagram as in Fig.~\ref{fig:sc} simulated for exact dynamics with ${J=0.04}$\,eV and harmonic potential.
    Changing $m_0$ tunes the system between topological bands (top, $m_0=1.4$\,eV) and trivial bands (bottom, $m_0=2.4$\,eV).
    The exciton ground state corresponds to the electron and hole both occupying the upper band of their respective copies of $H^\text{FB}_\text{BHZ}$, with the remaining parameters the same as in Fig.~\ref{fig:sc}.
    The white dashed curve again corresponds to the semiclassical boundary between the harmonic (dark blue) and Bloch (red) oscillation regimes.
    Large $\kappa$ corresponds to the exciton ground state being a wide wavepacket in relative momentum space.
    As a result, the group velocity is close to zero which suppresses the effect of the restoring force.
    When the band is topological, there is still a net transverse drift from average Berry curvature in this regime, in contrast to the semiclassical case in Fig.~\ref{fig:sc}.  
    }
\label{fig:exactdynamics}
\end{figure}

We simulate the exact dynamics of the exciton for a four band model with the same Berry curvature profiles and electron and hole dispersion as for the semiclassical numerics.  
We consider the Hamiltonian
\begin{align}
H_{\mathbf{K}}=&H^{\text{FB}}_{\text{BHZ},\mathbf{K}  }\otimes\mathds{1}_h+\mathds{1}_e\otimes H^{\text{FB}}_{\text{BHZ},\mathbf{K} } \notag 
\\ &\quad +\sum_{\mathbf{r}}V(\mathbf{r})\mathds{1}_e\otimes\mathds{1}_h\otimes\left|\mathbf{r}\right>\left<\mathbf{r}\right|
\label{HBHZcomfb}
\end{align}
where $H_{\text{BHZ},\mathbf{K}}^{\text{FB}}$ is a tight-binding Hamiltonian  obtained from the partial Fourier transform (performed in $k$ space) of the band-flattened BHZ Hamiltonian in Eq.~\eqref{eq:H-FB}, $\mathds{1}_{e/h}$ is the identity matrix for the electron/hole Hilbert space, and $V(\mathbf{r})$ is the potential modeling the interaction (either harmonic or Coulombic).
Note that as the interaction only depends on the relative coordinate, the Hamiltonian decouples into different $\mathbf{K}$ sectors.  
We initialize the system in the state 
\begin{align}
    \ket{\psi_0} = \sum_{\mathbf{K}} \omega(\mathbf{K}) \ket{\Phi_0(\mathbf{K})}
\end{align}
where $\omega(\mathbf{K})$ is a narrow Gaussian envelope and $\ket{\Phi_0(\mathbf{K})}$ is the ground state of the Hamiltonian projected into the exciton Hilbert space
\begin{align}
H^{\text{ex.}}_{\bf K}&= \hat{P}_e\,\hat{P}_h \, H_{\mathbf{K}}\,\hat{P}_h\,\hat{P}_e.
\end{align}
The operators $\hat{P}_{e/h}$ project onto the upper band of $H_\text{BHZ}^{\text{FB}},$ thereby ensuring the electron remains in the conduction band and the hole in the valence band.  
Our simulations use a real-space tight-binding approximation.
Further details of the numerics are given in Appendix~\ref{app:simulation}.

For a harmonic potential, Fig.~\ref{fig:exactdynamics} plots the average COM motion of the exciton over a Bloch cycle when the BHZ parameters are chosen such that the bands are topological (top panel) and trivial (bottom panel).
The former corresponds to $m_0=1.4$\,eV while the latter corresponds to $m_0=2.4$\,eV.
The remaining parameters are the same as in Fig.~\ref{fig:sc}.
The dashed curves again indicate the semiclassical boundary in Eq.~\eqref{eq:sm-limit}.
Just as was seen for semiclassical simulations of the harmonic potential, the plots remain the same when $J$, $E$, and $\kappa$ are scaled by the same factor.

For non-trivial Chern number, we observe a large transverse drift in the COM position throughout the semiclassical Bloch oscillation regime.
Additionally, we also observe the large $\kappa$ regime discussed in the previous section.
The latter has a smaller anomalous drift compared to the semiclassical regime.
There is no transverse drift when we choose our band projection such that the electron and hole bands have opposite Berry curvature.

\begin{figure}
	\includegraphics[scale=0.5]{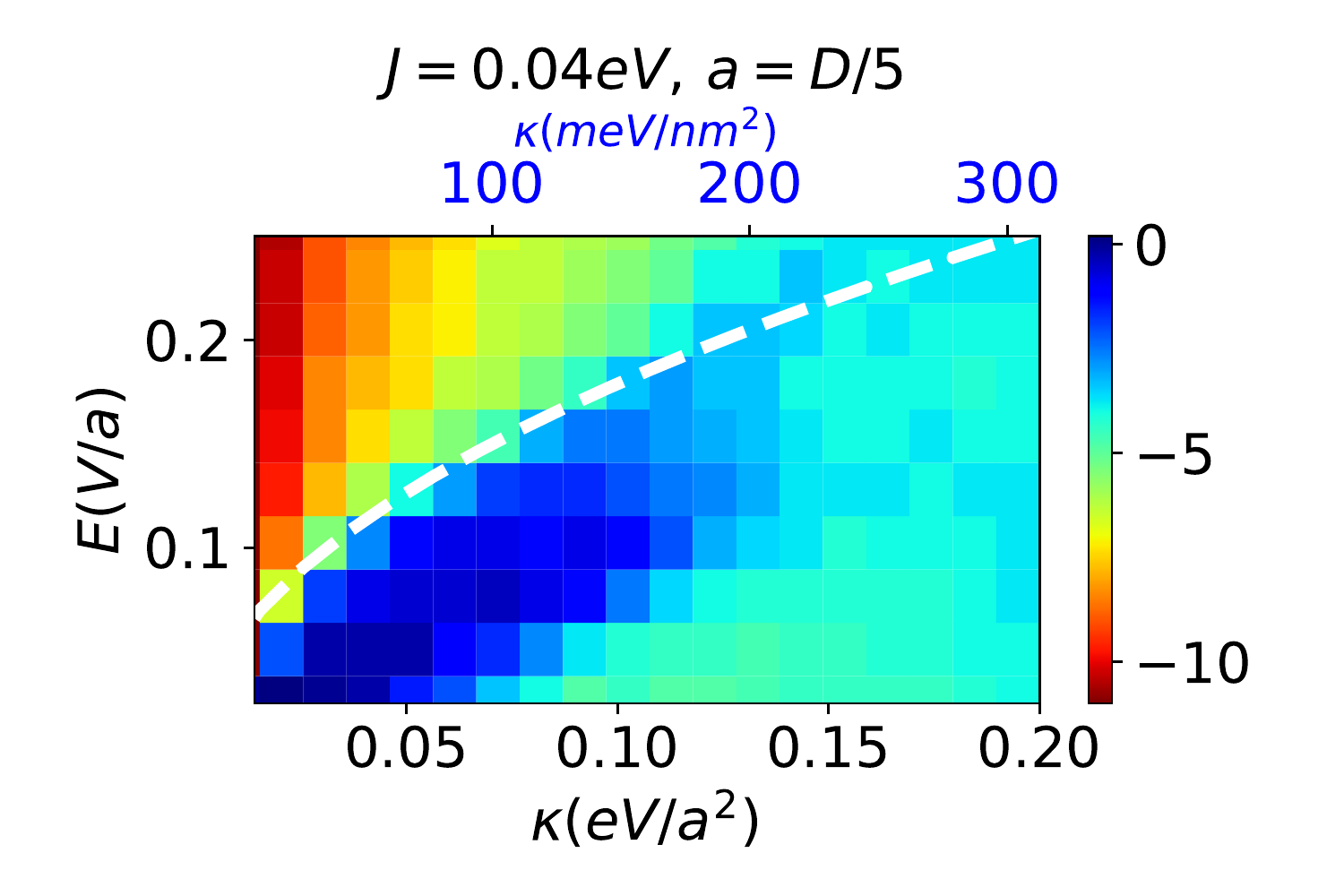}
	\includegraphics[scale=0.5]{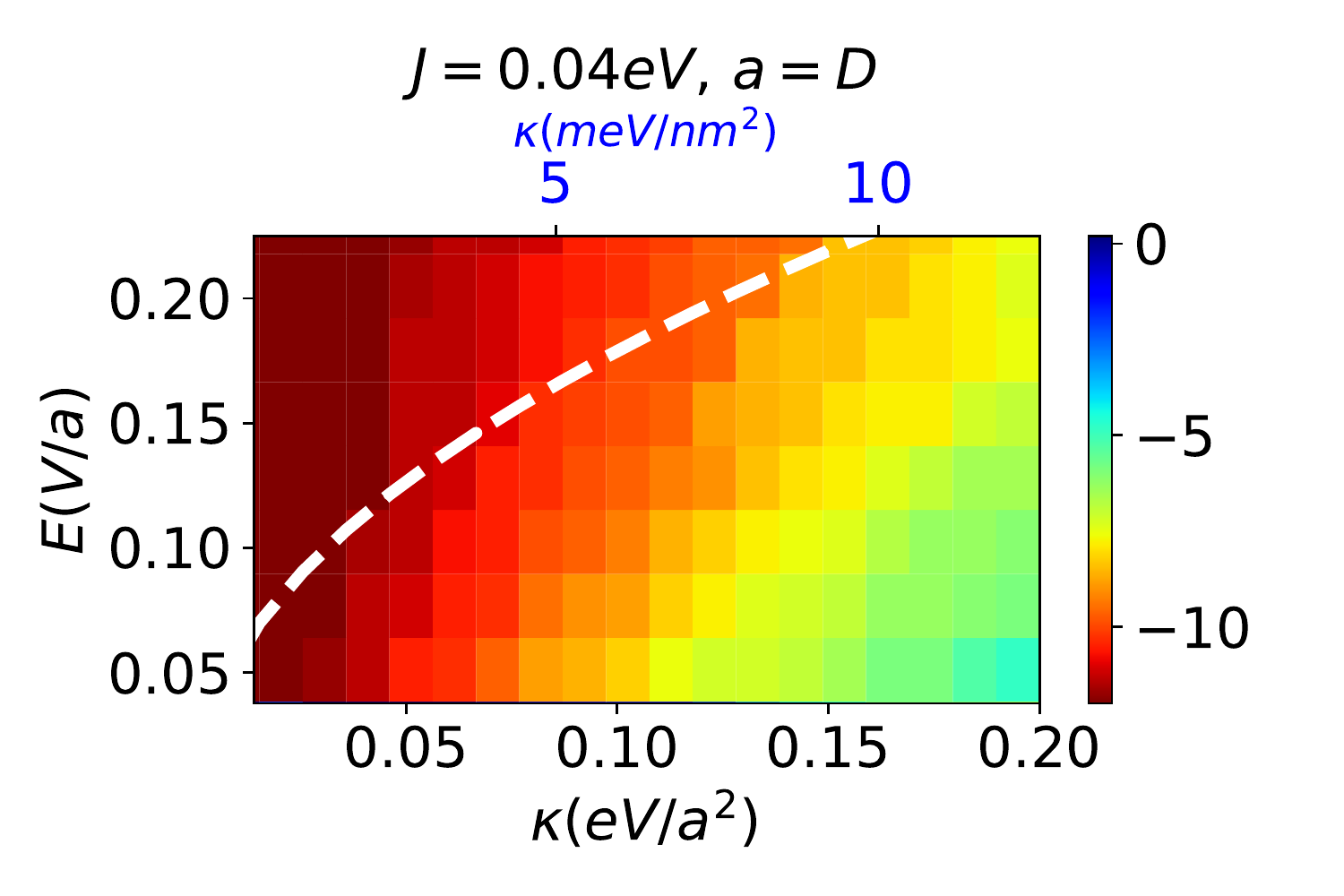}
\caption{{\it Exact dynamics for Coulombic potential: topological bands}.
    Same phase diagram as in Fig.~\ref{fig:sc} simulated for exact dynamics with ${J=0.04}$\,eV and Coulombic potential for ${a= D/5}$ (top) and ${a=D}$ (bottom).
    The exciton ground state corresponds to both electron and hole occupying the upper band of Eq.~\eqref{eq:H-BHZ} with the same parameters as in Fig.~\ref{fig:sc}.  
}
\label{fig:exactdynamicsCoulombic}
\end{figure}

The bottom panel of Fig.~\ref{fig:exactdynamics} demonstrates that trivial electronic bands can still support an anomalous exciton drift, albeit of reduced magnitude.
We see the effect only exists for the semiclassical Bloch oscillation regime corresponding to small $\kappa$ and large $E$.
This can be understood as a consequence of large $\kappa$ binding the ground state wavefunction more tightly in relative real space: as a result, the wavefunction spreads in relative momentum space, and thus experiences an averaged Berry curvature.
The averaged Berry curvature approaches the Chern number $\mathcal{C}$ over the area of the Brillouin zone, and thus becomes vanishingly small when $\mathcal{C}=0$.
Correspondingly, trivial bands do not support a large $\kappa$ regime of anomalous exciton transport.

\begin{figure}
	\includegraphics[scale=0.5]{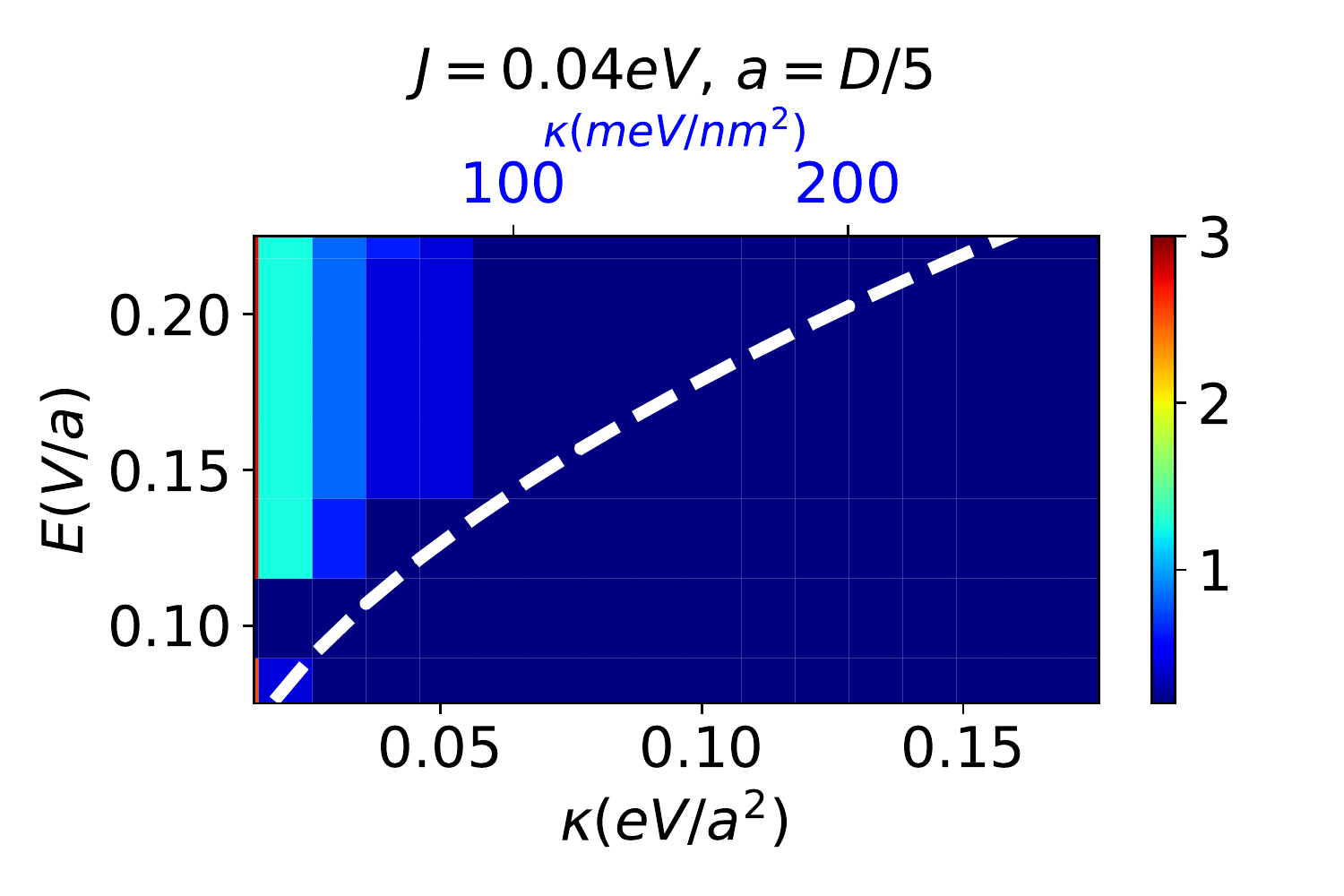}
	\includegraphics[scale=0.5]{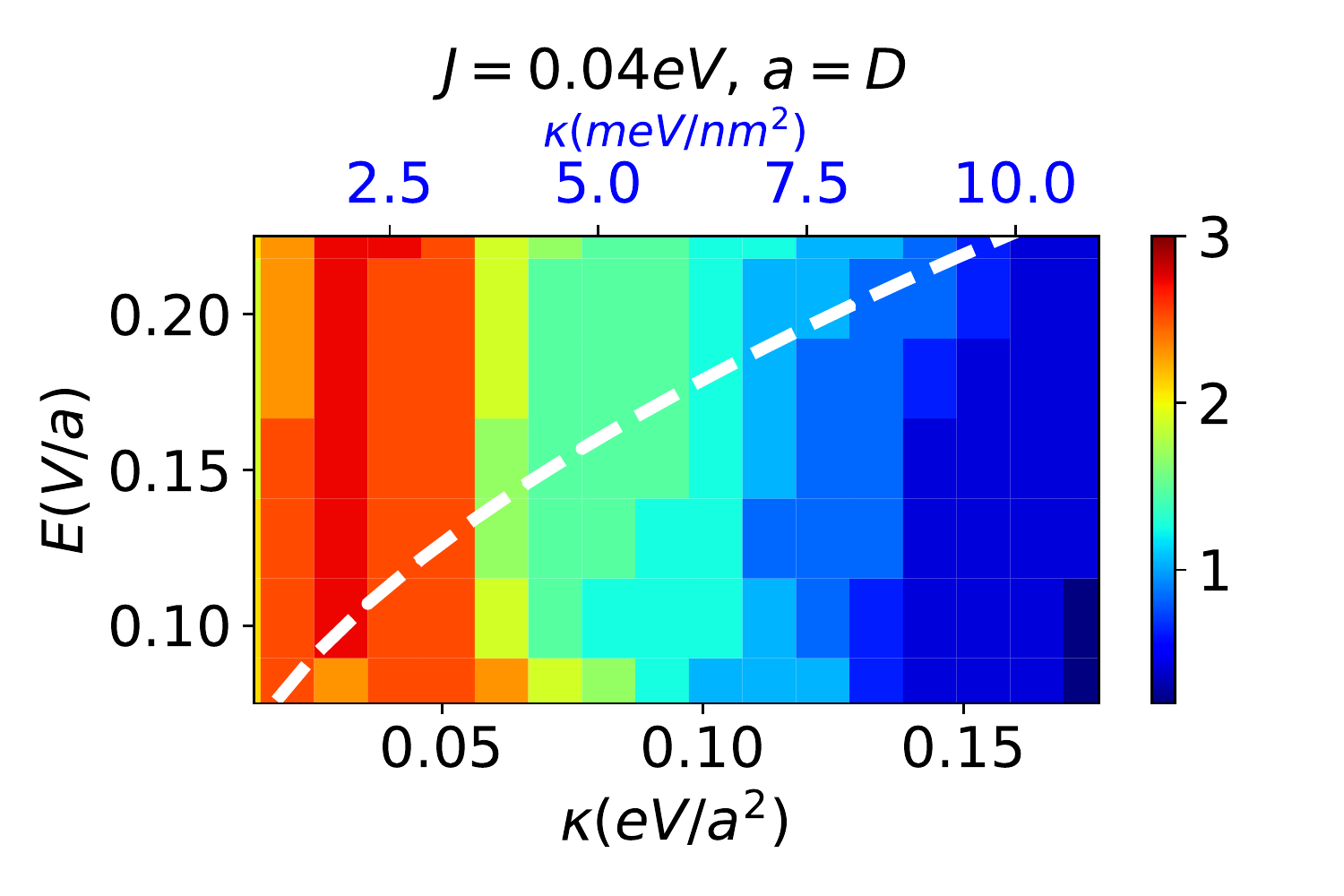}
\caption{{\it Exact dynamics for Coulombic potential: trivial bands}.
    Same phase diagram as in Fig.~\ref{fig:sc} simulated for exact dynamics with $J=0.04$\,eV and Coulombic potential for $a=D/5$ (top) and $a=D$ (bottom).
    The exciton ground state corresponds to both electron and hole occupying the upper band of Eq.~\eqref{eq:H-BHZ} with the same parameters as in Fig.~\ref{fig:sc} except for $m_0=2.4$\,eV,  corresponding to the trivial regime.
}
\label{fig:exactdynamicsCoulombictrivial}
\end{figure}

Figures~\ref{fig:exactdynamicsCoulombic} and \ref{fig:exactdynamicsCoulombictrivial} plot the anomalous exciton drift for the case of Coulomb interaction with topological and trivial bands, respectively.  
We take the same parameters as for Fig.~\ref{fig:exactdynamics}.
As expected, when $a\ll D$ (top panels), we see good agreement with Fig.~\ref{fig:exactdynamics}, including the `large $\kappa$' Bloch oscillation regime for the case of topological bands.
When $a=D$, we again see an increased Bloch oscillation regime.
The bottom panel of Fig.~\ref{fig:exactdynamicsCoulombic} can be understood analogously to the semiclassical simulation with Coulomb interactions in Fig.~\ref{fig:scCoulombic}.
Note that for fixed $D$, the bottom panel corresponds to a smaller range of $E$ and $\kappa$ compared to the top panel (see top axis in blue).

\section{Candidate Physical Systems}\label{sec:candidate-systems}

In the previous sections, we demonstrated regimes of anomalous exciton transport in response to a uniform in-plane electric field.
We now discuss additional complications beyond the scope of the models considered.
We posit that transition metal dichalcogenide (TMD) heterobilayers are potential platforms for hosting this effect due to their ability to support intervalley, interlayer excitons with large binding energies and long lifetimes.
Moir\'e TMDs are especially intriguing given the presence of flat, topological bands.
However, the large moir\'e lattice period compared to the exciton Bohr radius, as well as the presence of additional bands, add complications beyond the scope of the current analysis.
At the end of this section, we describe potential measurement schemes for observing anomalous exciton transport.

\subsection{Physical constraints}

The toy models considered earlier demonstrate that in principle an exciton can move in response to a uniform in-plane electric field when the constituent electron and hole undergo Bloch oscillations.
We now discuss additional physical constraints not captured by these models.
We reinsert factors of $\hbar$ throughout this section for ease of conversion to physical units.

First, if the electric field is sufficiently large, the gain in potential energy from spatially separating the electron and hole can overcome the binding energy $\varepsilon_B$.
When this occurs, the electron and hole dissociate into two freely moving particles and there is no well-defined exciton.  
For the effect considered here, the electron and hole undergo Bloch oscillations and thus their maximum spatial separation is bounded by the Bloch amplitude $x_\text{Bloch}$ in Eq.~\eqref{eq:x-Bloch}.
Therefore, provided the bandwidth does not exceed the binding energy, the exciton remains well-defined throughout the Bloch oscillation regime, i.e. we require
\begin{align}\label{eq:EB-bound}
    \varepsilon_B &> e E \,x_\text{Bloch} \sim 2J. 
\end{align}

When a system has multiple electronic bands, we also need to consider the possibility of Landau-Zener transitions.
For instance, an electron transitioning to a higher band effectively increases the bandwidth, potentially allowing the electron and hole to reach their equilibrium position, thereby transitioning to the harmonic oscillator regime (no anomalous COM drift).
For a Landau-Zener Hamiltonian $H_\text{LZ} = ct \sigma_z + \lambda \sigma_x$, the transition probability is given by $p=\text{exp}\left\{- \pi \lambda^2/\hbar c \right\}$.
Minimizing these transitions thus amounts to finding a regime where the sweep rate $c$ satisfies $\hbar c\ll \pi \lambda^2.$ 
In the case of an electron transitioning out of the conduction band, $\lambda$ corresponds to the minigap at the Brillouin zone boundary.
We can roughly approximate the sweep rate $c$ as the linearized slope of the band $J/(\pi/a)$ multiplied by $\hbar \dot{k}_e = eE$ (neglecting interactions), so that $\hbar c=(J a e E)/\pi$.
Landau-Zener transitions can then be neglected, provided that 
\begin{align}\label{eq:LZ-bound}
    E \ll E_\text{max}=\frac{\pi^2 \lambda^2}{J a e}.  
\end{align}
As such, flatter bands and larger minigaps can sustain a larger electric field, and thus a stronger effect.
Note that Eq.~\eqref{eq:LZ-bound} competes with the lower bound of the semiclassical regime given in Eq.~\eqref{eq:sm-limit}, but not with the large $\kappa$ regime identified in our simulations.

Additionally, when evaluating the attractiveness of any candidate physical system, we must further consider the time scales of the exciton.
Clearly, the exciton lifetime must be sufficiently long that the anomalous drift can be observed.
At a minimum, this requires the exciton lifetime exceeding the Bloch oscillation period.
Moreover, if the exciton relaxes to its equilibrium position, e.g. through phonon scattering, the anomalous velocity will vanish.
Provided the energy separation between the excited exciton (undergoing anomalous drift) and the (stationary) exciton ground state is less than the optical phonon band gap, we only need to consider acoustic phonon scattering.  
Assuming the bands are flat enough that the electron and hole group velocities are slower than the speed of sound, such scattering only occurs when the exciton hits an impurity and should therefore be negligible for sufficiently clean systems.  

Finally, the anomalous velocity grows linearly with $\mathbf{\Omega}^c_e(\mathbf{k}_e)-\mathbf{\Omega}^v_h(\mathbf{k}_h)$, thus a system that hosts intervalley excitons and topological bands will have a stronger response.
We emphasize that topological bands are not a prerequisite (see Fig.~\ref{fig:exactdynamics} and \ref{fig:exactdynamicsCoulombictrivial}), but will make the effect more visible.

\subsection{TMD heterobilayers}

TMDs are an excellent platform to study Berry curvature effects on excitonic properties: excitons in these materials have large binding energies and dominate the optical responses of the system.
In particular, we posit that TMD heterobilayers are an attractive platform to observe the anomalous excitonic drift studied in this paper.

One of the key requirements of the anomalous excitonic drift is the formation of intervalley excitons so that the electron and hole bands have opposite Berry curvature.
In a TMD monolayer, such an exciton requires a large COM momentum and thus is optically dark.
However, a TMD heterobilayer with a type-II band alignment (e.g., MoX$_2$/WX$_2$) supports excitons whose electron and hole are localized in different layers. 
When the two layers are twisted by an angle $\theta\approx60^\circ$ (Fig. 3(g) in Ref.\cite{Yu15}), the system can support an intervalley exciton with close to zero COM momentum.
There are two distinct benefits: (1) such an exciton can be optically bright and as such can be easily excited and detected, and (2) the spatial separation of the electron and hole enhances the exciton lifetime to anywhere from hundreds of nanoseconds to a few microseconds~\cite{Rivera2015, Rivera18}.

TMD heterobilayers have a slight lattice mismatch.
When the layers are closely aligned, a moir\'e potential forms with amplitude up to ${\sim 150}$\,meV~\cite{Rivera18}  and lattice period up to ${\sim 20}$\,nm~\cite{Wu18}.
As a result, the electronic bands flatten to a bandwidth ${\sim10-50}$\,meV~\cite{Ruiz-Tijerina19}, which can be adjusted further by changing the twist angle.
The resulting interlayer excitons retain a large binding energy  ${\sim 100-200}$\,meV and a Bohr radius $\sim2$\,nm~\cite{Ruiz-Tijerina19}.
At first glance, moir\'e TMDs seem especially promising for observing anomalous excitonic drift due to the flatter bands and similar binding energy making the Bloch oscillation regime more accessible.  
We might further hope that the possibility of topological moir\'e bands~\cite{Sie15,Wu17, Wu18, Kwan20} and the larger moir\'e lattice period would result in a more pronounced exciton anomalous velocity.
We note there are two features that complicate interpretation of our numerics for moir\'e TMDs.
First, the exciton's Bohr radius is significantly smaller than the moir\'e lattice period; in our simulations this corresponds to the large $\kappa$ region of phase space only (for which the ground state wavefunction extent is less than a lattice constant).
Second, our assumption that electron and hole occupy a single band may not apply given the reduced size of the moir\'e Brillouin zone.
Survival of the anomalous excitonic drift in moir\'e TMDs remains an interesting open question we plan to investigate in a future work.

A back of the envelope estimate suggests the parameters of TMD heterobilayers are compatible with the bounds identified in the previous section.
The binding energy $\sim100-200$\,meV easily exceeds the typical bandwidth $\sim10-50$\,meV, satisfying the necessary condition in Eq.~\eqref{eq:EB-bound} to avoid exciton ionization.
The upper bound on the electric field in Eq.~\eqref{eq:LZ-bound} from Landau Zener transitions is compatible with the lower bound in Eq.~\eqref{eq:sm-limit} from the semiclassical Bloch oscillation regime.
For instance, the anti-parallel configuration of MoSe$_2$/WS$_2$ has a bandwidth $J\sim5$\,meV, energy gap between lowest flat band to next moir\'e band $\lambda\sim 20$\,meV, and a lattice constant $a\sim 8$\,nm~\cite{Ruiz-Tijerina19,Ruiz-Tijerina20}, corresponding to $E_\text{max}\sim \pi^2 \lambda^2/(Jae) \sim 60$\,mV/nm. 
Taking interlayer separation $D\approx3$\,nm, dielectric constant $\epsilon\approx 4$, and interaction parameter $\kappa\approx 20$\,meV/nm$^2$~\cite{Berman17}, $E_\text{max}>E_\text{min}\sim 2\sqrt{2J\kappa}\approx 30$\,mV/nm.  
We further note that optical phonons in most TMDs have energies greater than $30$\,meV~\cite{Ovesen19,Sanchez11}; given that the energy gained by the exciton is on the order of the bandwidth $J$, the exciton cannot relax to its ground state by emitting a phonon.  
As noted previously, relaxation from acoustic phonon scattering can be neglected for sufficiently clean systems.

\subsection{Measurement schemes}

Lastly, we discuss possible measurements to observe the anomalous exciton drift in TMD heterobilayers.
As noted in the previous section, several TMD heterobilayers naturally support optically bright intervalley, interlayer excitons~\cite{Rivera18}.
Thus, we consider a situation where excitons are excited by illuminating one side of the sample, a uniform in-plane electric field $\mathbf{E}=E\mathbf{\hat{x}}$ is turned on, and we look for signatures of the excitons in the transverse direction.

The exciton trajectories can be directly observed using photoluminescence.
Polarization-resolved photoluminescence has been proposed~\cite{Yao08b,Yun-Mei15,Kuga08} and used~\cite{Onga17} to observe the excitonic Hall effect on the micron scale.
A similar approach could be used here, provided the anomalous drift survives sufficiently many Bloch cycles. 
A photoluminescence measurement in the transverse direction from where the excitons are initially excited should have a stronger response than the same measurement performed in the direction parallel to the electric field.  

An alternative approach is to separately contact and measure the current in the TMD layers.
For a TMD heterobilayer with type-II band alignment, all interlayer excitons have electrons localized to one layer, and holes to the other.
As such, the anomalous exciton drift should manifest as a current in the transverse direction (positive for one layer, negative for the other.
Separately contacting the layers requires an insulating layer inserted between the TMDs so as not to short-circuit the sample.
Interlayer excitons have been observed in TMD monolayers separated by hBN~\cite{Fang14,Rivera18,Calman18}.
This approach is analogous to a Coulomb drag measurement, in which a voltage is applied in one layer and the current is measured in the other.
Coulomb drag has previously been used to measure spatially indirect exciton transport in bilayer 2DEGs~\cite{Nandi12}.

Other potential measurement schemes could utilize the out-of-plane dipole moment of the interlayer excitons participating in the effect, or the thermal gradient resulting from exciton transport across the system.
The former would require measuring the dipole density to detect that excitons excited on one edge of the sample had traveled in the transverse direction.
Both such measurements would likely require a high density of excitons to be observable, as could be provided by an exciton condensate.

\section{Discussion and Outlook} \label{sec:discussion}

In this work, we have studied anomalous exciton drift in response to a uniform in-plane electric field.
We have demonstrated this effect semiclassically for intervalley excitons when the electron and hole bands have finite Berry curvature.
We have further simulated a toy model exhibiting this effect for a range of electric field and interaction strengths.
Our numerics indicate a Bloch oscillation regime not predicted by semiclassics, which we can analytically understand through a simple 1D model.
We have postulated that TMD heterobilayers are an attractive candidate system for observing anomalous exciton transport.

Previous works have also considered anomalous exciton transport resulting from finite Berry curvature when the exciton center of mass experiences a net force~\cite{Yao08b,Kovalev19,Kwan20,Cao20}.
As we were completing this manuscript, Ref.~\onlinecite{Cao20} by Cao, Fertig, and Brey, was posted.
Cao et al. propose a similar anomalous effect can arise from a COM momentum-dependent dipole curvature of the exciton ground state, originating from the geometry of the exciton ground state.
They primarily consider excitons in a magnetic field, with the exception of excitons in bilayer graphene (Sec. IV ibid.), where an asymmetry of the two layers is required for a non-vanishing effect. 
In contrast, the anomalous exciton transport established here is a dynamical effect at zero magnetic field, that cannot be accounted for without considering the internal exciton dynamics and binding interaction. 
Nonetheless, the underlying origin of both proposals is related, particularly in the small field limit. 
We leave a detailed comparison of our results with Ref.~\onlinecite{Cao20} to future work. 

Lastly, we note that moir\'e TMDs remain an interesting potential platform for the anomalous exciton drift due to the flat bands, enhanced Berry curvature, and large lattice spacing.
We emphasize that additional care is needed to apply our results to these systems given our assumption that electron and hole each occupy a single band.
Potentially, more complicated TMD heterostructures might also provide a platform for observing the effect, for instance a pair of moir\'e TMD bilayers separated by insulating hBN layers.

\acknowledgements
We are grateful to Felix von Oppen and Michael Fogler for stimulating conversations.
This work was supported by the Institute of Quantum Information and Matter, an NSF Frontier center funded by the Gordon and Betty Moore Foundation, the Packard Foundation, and the Simons Foundation.
SC and GR thank the U.S. Department of Energy, Office of Science, Basic Energy Sciences under Award de-sc0019166.
GR is also grateful to the NSF DMR grant number 1839271.
NSF and DOE supported GR's time commitment to the project in equal shares.  
CK acknowledges support from the Walter Burke Institute for Theoretical Physics at Caltech.


\appendix

\section{Relation between Berry curvature of electron and hole in a given band}
\label{app:derivation}

The Berry curvature $\mathbf{\Omega}_\alpha$ and Berry connection $\mathbf{A}_\alpha$ for a band $\alpha$ can be defined as 
\begin{align}
\Omega_\alpha(\mathbf{k}) &= \nabla_{\mathbf{k}} \times \mathbf{A}_\alpha (\mathbf{k})
\\ \mathbf{A}_\alpha(\mathbf{k}) &=  \bra{u_\alpha}i\nabla_{\mathbf{k}}\ket{u_\alpha},
\end{align}
where $\ket{u_\alpha}$ is the Bloch state for band $\alpha$.  
We can write the Berry connection in terms of the Bloch wavefunctions using 
\begin{equation}
\begin{split}
\mathbf{A}_\alpha(\mathbf{k}) &= \int d\mathbf{r} \bra{u_\alpha}\mathbf{r}\rangle i\nabla_{\mathbf{k}}\bra{\mathbf{r}}u_\alpha\rangle
\\&= i \int d\mathbf{r} \left( u_{\alpha,\mathbf{k}}(\mathbf{r})\right)^*\nabla_{\mathbf{k}} u_{\alpha,\mathbf{k}}(\mathbf{r}).
\end{split}
\end{equation}
In order to understand the connection between the Berry curvature for a hole in band $\alpha$ compared to the Berry curvature for an electron in the same band, we can assume that the creation operator for a hole in band $\alpha$ at momentum $\mathbf{k}$ is equal to the annihilation operator for an electron in band $\alpha$ at momentum $-\mathbf{k}$:
\begin{align}
d^\dagger_{\mathbf{k},\alpha} &= c_{-\mathbf{k},\alpha},
\end{align}
where $d$ is for the hole and $c$ is for the electron. 
In real space we have
\begin{align}
c_\alpha^\dagger(\mathbf{r}) &= d_\alpha(\mathbf{r}).
\end{align}
Therefore, we see 
\begin{align}
&c_{\mathbf{k},\alpha}^\dagger = \int d\mathbf{r} e^{i\mathbf{k}\cdot\mathbf{r}}u_{\alpha,\mathbf{k}}^e(\mathbf{r}) c_\alpha^\dagger(\mathbf{r})
\\ &d_{-\mathbf{k},\alpha} = \int d\mathbf{r} e^{i\mathbf{k}\cdot\mathbf{r}} \left(u_{\alpha,-\mathbf{k}}^h(\mathbf{r})\right)^* d_\alpha(\mathbf{r})
\\ &\Rightarrow u_{\alpha,\mathbf{k}}^e (\mathbf{r}) = \left( u_{\alpha,\mathbf{k}}^h(\mathbf{r})\right)^*.  
\end{align}
Using the above equations, the Berry connection for the hole can be related to the Berry connection of the electron by
\begin{align}
\mathbf{A}_\alpha^h (\mathbf{k}) &= i \int d\mathbf{r}  \left( u_{\alpha,\mathbf{k}}^h(\mathbf{r})\right)^* \nabla_{\mathbf{k}} u_{\alpha,\mathbf{k}}^h(\mathbf{r}) 
\\ &=  i \int d\mathbf{r}  u_{\alpha,-\mathbf{k}}^e(\mathbf{r}) \nabla_{\mathbf{k}} \left(u_{\alpha,-\mathbf{k}}^e(\mathbf{r}) \right)^*
\\ &=  -i \int d\mathbf{r}  \left( u_{\alpha,-\mathbf{k}}^e(\mathbf{r}) \right)^*\nabla_{\mathbf{k}}u_{\alpha,-\mathbf{k}}^e(\mathbf{r})
\\ &=  i \int d\mathbf{r}  \left( u_{\alpha,-\mathbf{k}}^e(\mathbf{r}) \right)^*\nabla_{-\mathbf{k}}u_{\alpha,-\mathbf{k}}^e(\mathbf{r})
\\ &=  \mathbf{A}_\alpha^e(-\mathbf{k}).
\end{align}
It follows that the Berry curvatures are related by 
\begin{align}
\begin{split}
\mathbf{\Omega}_\alpha^h(\mathbf{k}) &= \nabla_{\mathbf{k}}\times \mathbf{A}_\alpha^h(\mathbf{k}) = \nabla_{\mathbf{k}} \times \mathbf{A}_\alpha^e(-\mathbf{k}) \\&= - \nabla_{-\mathbf{k}} \times \mathbf{A}_\alpha^e(-\mathbf{k}) = - \mathbf{\Omega}_\alpha^e(-\mathbf{k}).
\end{split}
\end{align}
Now, if the momentum of the created hole is $\mathbf{k}_h$, then the momentum of the electron that was removed is ${\mathbf{k}_e=-\mathbf{k}_h}$ and
\begin{align}\label{eq:ass-2}
\mathbf{\Omega}^h_\alpha(\mathbf{k}_h) &= -\mathbf{\Omega}_\alpha^e(\mathbf{k}_e).
\end{align}

\section{Intermediate semiclassical case}\label{app:intermediate}

In the main text, we considered the fine-tuned limit of equal Berry curvature for electron and hole, $
{\mathbf{\Omega}^c_e(\mathbf{k}) = \mathbf{\Omega}^v_e(\mathbf{k})}$, equivalently $
{\mathbf{\Omega}^c_e(\mathbf{k}) = -\mathbf{\Omega}^v_h(-\mathbf{k})}$ from Eq.~\eqref{eq:ass-2}.  
In this particular case, Berry curvature effects appear only in COM motion. 
Similarly, for ${\mathbf{\Omega}^c_e(\mathbf{k}) = \mathbf{\Omega}^v_e(-\mathbf{k})}$, Berry curvature only affects the relative motion. 
However, in type-II heterobilayers none of these conditions are satisfied exactly, and Berry curvature effects couple the COM and relative space equations of motion. 
For a direct momentum exciton, the relative and COM position evolve according to
\begin{align}
\dot{\mathbf{r}} &= \left( 2Ja\sin\left( k_xa\right)-\dot{k}_y\Delta\Omega^{vc}_e(\mathbf{k})\right)\,\mathbf{\hat{x}} \notag 
\\ &\quad+ \left(2Ja\sin\left( k_ya\right)+
\dot{k}_x\Delta\Omega^{vc}_e(\mathbf{k})\right)\, \label{eq:int-1}
\\ \dot{\mathbf{R}} &=2\dot{k}_x\Omega^\text{avg}_e(\mathbf{k})\, \mathbf{\hat{y}}-2\dot{k}_y\Omega^\text{avg}_e(\mathbf{k})\,\mathbf{\hat{x}} \label{eq:int-2}
\end{align}
where the difference and average Berry curvatures are defined by 
\begin{align}
   \Delta\Omega^{vc}_e(\mathbf{k}) &=\Omega^c_e(\mathbf{k})-\Omega^v_e(\mathbf{k}) \\ \Omega^\text{Avg}_e(\mathbf{k})&=\frac{1}{2}\left(\Omega^c_e(\mathbf{k})+\Omega^v_e(\mathbf{k})\right). 
\end{align}

\begin{figure}
    \centering
    \includegraphics[scale=0.5]{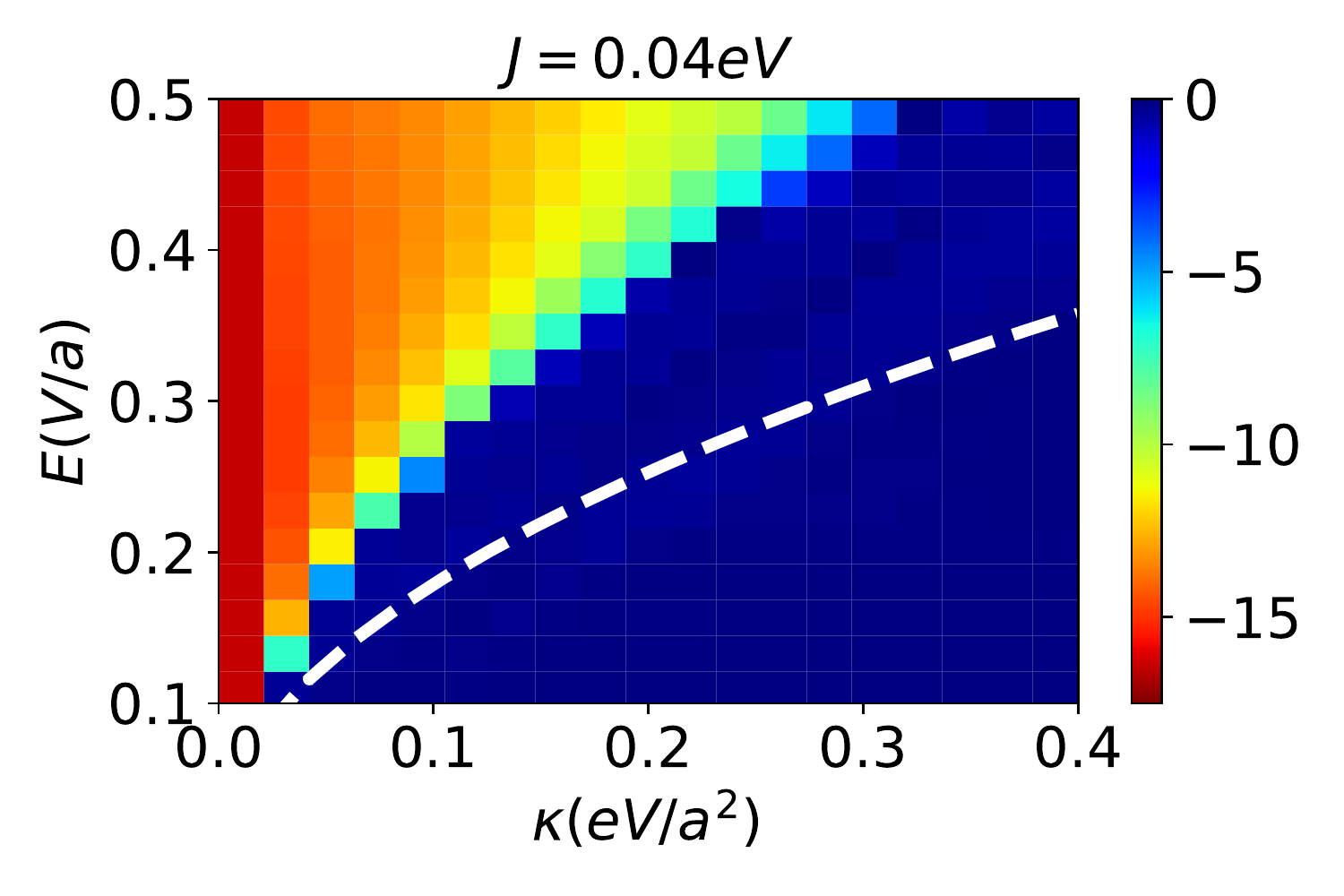}
    \includegraphics[scale=0.28]{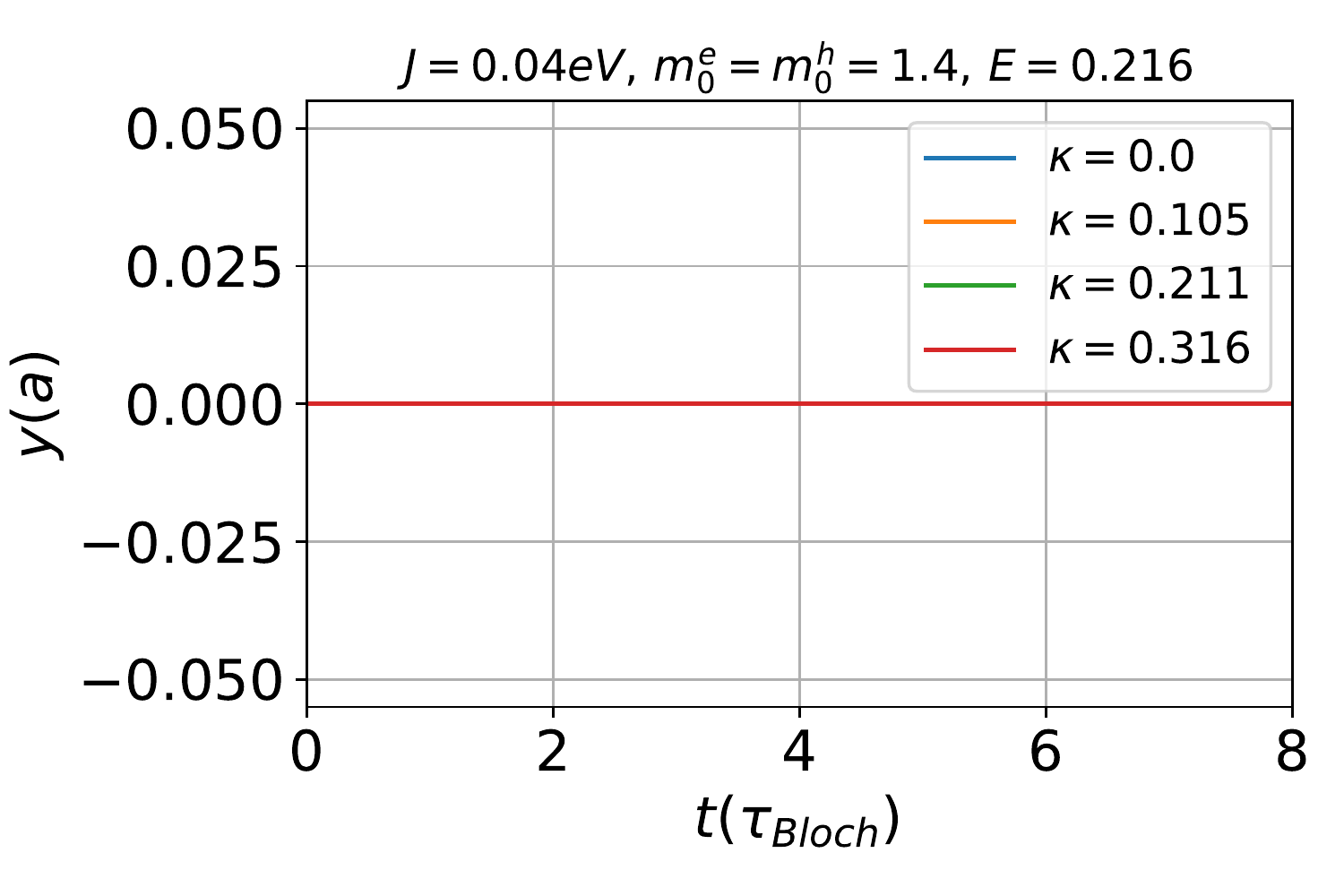}
    \includegraphics[scale=0.28]{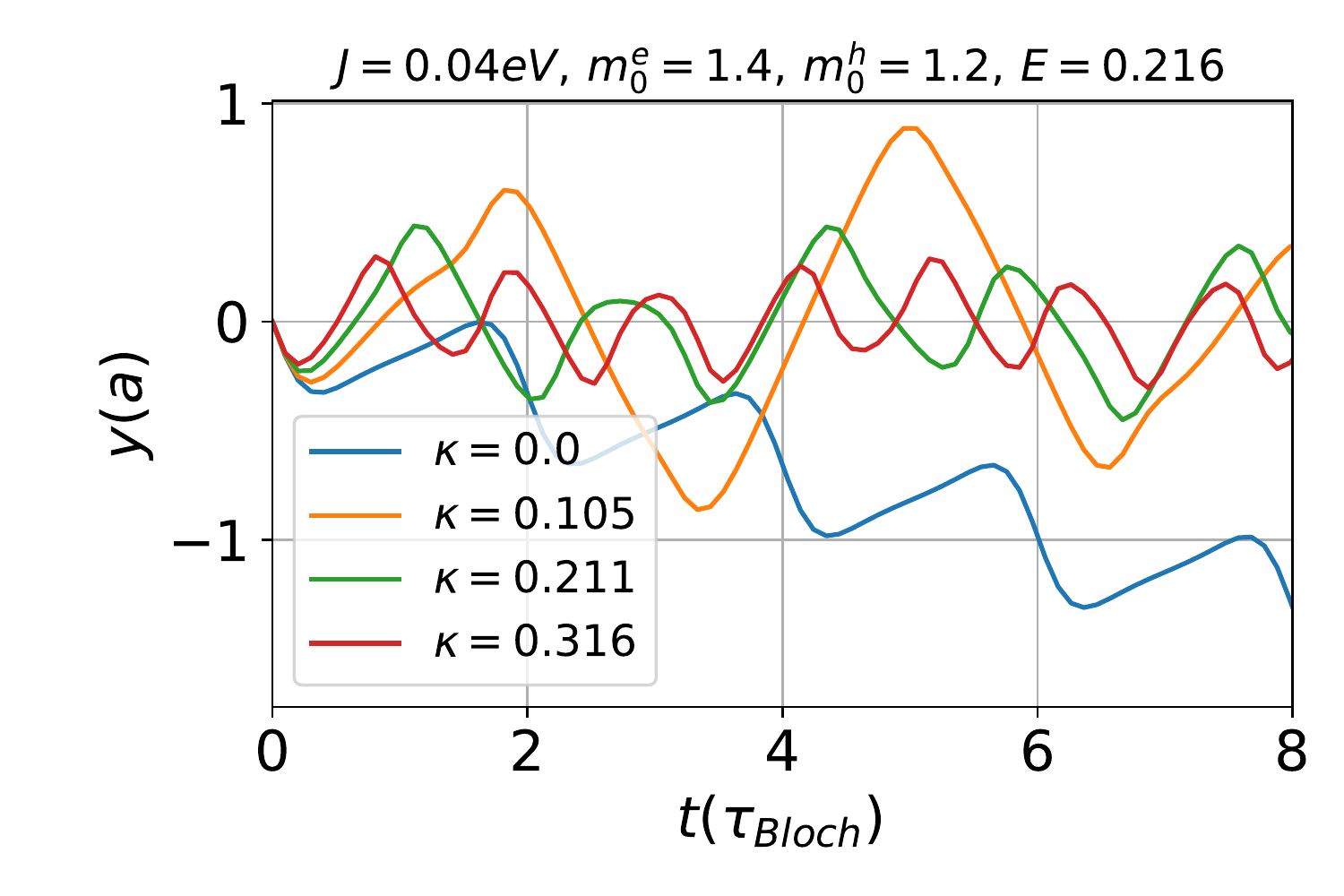}
    \caption{{\it Intermediate semiclassical dynamics.} 
    The top panel plots average $Y$ per Bloch cycle when $\Delta\Omega^{vc}_e(\mathbf{k})\neq 0.$
    To give the electron and hole bands slightly different Berry curvature we use $m_0^e=1.4$\,eV and $m_0^h=1.2$\,eV.
    The other parameters are the same as in Fig.~\ref{fig:sc}.
    The dashed white curve is again the semiclassical boundary $E=2\sqrt{2J\kappa}$ shown in Figs.~\ref{fig:sc}-\ref{fig:exactdynamicsCoulombic}; Berry curvature couples $Y$ and $y$, thereby reducing the Bloch oscillation regime compared to the symmetric case considered in Fig.~\ref{fig:sc}.
    In the bottom panel, we compare transverse drift in $y$ as a function of time for equal (left) and different (right) electron and hole Berry curvatures. 
    }
    \label{fig:intermediate}
\end{figure}
Bloch oscillations are obfuscated in the relative motion when $\mathbf{R}$ and $\mathbf{r}$ are coupled.  
We continue to define the Bloch period as $\tau_\text{Bloch}=2\pi/(a e E)$.

We plot the effect of $\Delta\Omega^{vc}_e(\mathbf{k})\neq 0$ in Fig.~\ref{fig:intermediate}.  
The top panel shows the average transverse COM drift per Bloch cycle in $E$ versus $\kappa$ space.
The magnitude of the transverse drift is less than when the Berry curvatures of the electron and hole bands are equal (Fig.~\ref{fig:sc}), thus a system that approaches particle-hole symmetry should have a stronger anomalous exciton response.
We note the transition between harmonic and Bloch oscillation regimes is affected by the fact that relative and COM motion are now coupled.  
The bottom panel plots the relative transverse motion when $\Delta\Omega^{vc}_e(\mathbf{k})=0$ (right) and $\Delta\Omega^{vc}_e(\mathbf{k})\neq 0$ (left).
As predicted by Eqs.~\eqref{eq:int-1} and \eqref{eq:int-2}, the former corresponds to no Berry curvature effects on $y$, while the latter corresponds to $y$ and $Y$ being coupled.

\section{Comparison of semiclassical approaches}
\label{app:scdynamics}

We review different semiclassical approaches used to study the dynamics of electrons and excitons in Bloch bands. 
We contrast them with the semiclassical and exact dynamics approach used in the main text.

\subsection{Semiclassical description for non-interacting electron wavepacket in a Bloch band}

In this subsection, we review the semiclassical dynamics of a non-interacting electron wavepacket in a Bloch band. 
We closely follow the approach presented in Ref.~\onlinecite{Sundaram99}.
Consider a wavepacket in $k$-space described by the wavefunction
\begin{equation}
\ket{\Psi(t)}=\int d\mathbf{k}\, a(\mathbf{k},t)\ket{\psi_n(\mathbf{k})},
\end{equation}
where $\left|\psi_n(\mathbf{k})\right>=\sum_{\mathbf{r}} e^{i\mathbf{k}\cdot\mathbf{r}}\ket{u_n(\mathbf{k})} \otimes \ket{\mathbf{r}}$ are Bloch wavefunctions of $n^{th}$ eigenstates and $|a(\mathbf{k},t)|^2$ is centered around the point
\begin{equation}
\mathbf{k}_c=\int d\mathbf{k}( \mathbf{k} |a(\mathbf{k},t)|^2).
\end{equation}
We express
\begin{equation}
a(\mathbf{k},t)=|a(\mathbf{k},t)|e^{-i\gamma(\mathbf{k},t)}.
\end{equation}
The center of wavepacket in real space is
\begin{align}
\mathbf{r}_c&=\bra{\Psi}\mathbf{r}\ket{\Psi}
\\ &=\nabla_{\mathbf{k}}\gamma(\mathbf{k},t)|_{\mathbf{k}=\mathbf{k}_c}+\left<u(\mathbf{k})|i\nabla_{\mathbf{k}}|u(\mathbf{k})\right>|_{\mathbf{k}=\mathbf{k}_c}
\\ &=\nabla_{\mathbf{k}_c}\gamma(\mathbf{k}_c,t)+\left<u(\mathbf{k}_c)|i\nabla_{\mathbf{k}_c}|u(\mathbf{k}_c)\right>.
\end{align}

The dynamics of the mean position $\mathbf{r}_c$ and momentum $\mathbf{k}_c$ can be obtained using a time-dependent variational principle with the Lagrangian 
\begin{align}\label{eq:Lagrangian}
L=\bra{\Psi}i\frac{d}{dt}-H\ket{\Psi},
\end{align}
where $H=H_\text{Bloch}-e\mathbf{E}\cdot\mathbf{r}$. We have
\begin{align}
\left<\Psi|i\frac{d\Psi}{dt}\right>&=\int d\mathbf{k} |a(\mathbf{k},t)|^2 \notag \\&\times \bra{u(\mathbf{k},t)} e^{i\gamma(\mathbf{k},t)}\frac{d}{dt}(e^{-i\gamma(\mathbf{k},t)}\ket{u(\mathbf{k},t)})
\\&=\frac{\partial \gamma(\mathbf{k}_c,t)}{\partial t}+\left<u(\mathbf{k}_c,t)|i\frac{\partial}{\partial t}u(\mathbf{k}_c,t)\right>.
\end{align}
We can write 
\begin{equation}
\frac{\partial \gamma(\mathbf{k}_c,t)}{\partial t}=\frac{d\gamma(\mathbf{k}_c)}{dt}-\dot{\mathbf{k}_c}\cdot\frac{\partial \gamma(\mathbf{k}_c)}{\partial \mathbf{k}_c}
\end{equation}
and
\begin{equation}
\bra{\Psi}H\ket{\Psi}=\bra{\Psi}H_\text{Bloch}\ket{\Psi}-e\mathbf{E}\cdot\mathbf{r}_c=\mathcal{E}_\text{Bloch}-e\mathbf{E}\cdot\mathbf{r}_c.
\end{equation}
Now, the Lagrangian is
\begin{equation}
L=-\mathcal{E}(\mathbf{r}_c,\mathbf{k}_c)+\mathbf{k}_c\cdot\dot{\mathbf{r}}_c+\dot{\mathbf{k}}_c\cdot\left<u|i\frac{\partial u}{\partial \mathbf{k}_c}\right>+\left<u|i\frac{\partial u}{\partial t}\right>+\frac{d\gamma(\mathbf{k}_c,t)}{dt}.
\end{equation}
In the above, $\mathcal{E}(\mathbf{r}_c,\mathbf{k}_c)=\bra{\Psi}H_{\text{Bloch}}\ket{\Psi}-\mathbf{E}\cdot\mathbf{r}_c$, and we used  $\nabla_{\mathbf{k}_c}\gamma(\mathbf{k}_c,t)=\mathbf{r}_c-\bra{u(\mathbf{k}_c)}i\nabla_{\mathbf{k}_c}\ket{u(\mathbf{k}_c)}$.

This Lagrangian is a function of $\mathbf{r}_c$, $\dot{\mathbf{r}}_c$, $\mathbf{k}_c$, $\dot{\mathbf{k}}_c$, and $t$. If we assume  $\left<{u}|{i\frac{\partial u}{\partial t}}\right>=0$ (as is usually the case for adiabatic evolution and a translationally invariant system), the equations of motion for the wavepacket center are 
\begin{equation}
\frac{d}{dt}\left(\frac{\partial L}{\partial \dot{\mathbf{r}}_c}\right)-\frac{\partial L}{\partial \mathbf{r}_c}=0,\;\;\frac{d}{dt}\left(\frac{\partial L}{\partial \dot{\mathbf{k}}_c}\right)-\frac{\partial L}{\partial \mathbf{k}_c}=0 
\end{equation}
which implies
\begin{equation}
\frac{d \mathbf{k}_c}{dt}=-\frac{\partial \mathcal{E}(\mathbf{r}_c,\mathbf{k}_c)}{\partial \mathbf{r}_c},
\label{kc}
\end{equation}
\begin{widetext}
\begin{equation}
\begin{split}
\dot{x}_c=\frac{\partial\mathcal{E}}{\partial k_{xc}}+\frac{d}{dt}\left(\left<u|i\frac{\partial u}{\partial k_{xc}}\right>\right)-\dot{k}_{xc}\frac{\partial }{\partial k_{xc}}\left<u|i\frac{\partial u}{\partial k_{xc}}\right>-\dot{k}_{yc}\frac{\partial }{\partial k_{xc}}\left<u|i\frac{\partial u}{\partial k_{yc}}\right>
\end{split}
\end{equation}

\begin{equation}
\begin{split}
\dot{x}_c=\frac{\partial\mathcal{E}}{\partial k_{xc}}+\dot{k_{yc}}\frac{\partial}{\partial k_{yc}}\left(\left<u|i\frac{\partial u}{\partial k_{xc}}\right>\right)-\dot{k}_{yc}\frac{\partial }{\partial k_{xc}}\left(\left<u|i\frac{\partial u}{\partial k_{yc}}\right>\right)=\frac{\partial\mathcal{E}}{\partial k_{xc}}+\dot{k}_{yc}\left(\frac{\partial A_x}{\partial k_{yc}}-\frac{\partial A_y}{\partial k_{xc}}\right).
\end{split}
\label{vxc}
\end{equation}
In the above, $\mathbf{A}$ is the Berry-connection and  we use the fact that 
\begin{align}
\frac{d}{dt}\left(\left<u|i\frac{\partial u}{\partial k_{xc}}\right>\right)=\dot{k_{yc}}\frac{\partial}{\partial k_{yc}}\left(\left<u|i\frac{\partial u}{\partial k_{xc}}\right>\right)+\dot{k_{xc}}\frac{\partial}{\partial k_{xc}}\left(\left<u|i\frac{\partial u}{\partial k_{xc}}\right>\right)
\end{align}
as we already assumed $\bra{u}\ket{i\frac{\partial u}{\partial t}}=0$.
Similarly,
\begin{equation}
\begin{split}
\dot{y}_c=\frac{\partial\mathcal{E}}{\partial k_{yc}}+\dot{k_{xc}}\frac{\partial}{\partial k_{xc}}\left(\left<u|i\frac{\partial u}{\partial k_{yc}}\right>\right)-\dot{k}_{xc}\frac{\partial }{\partial k_{yc}}\left(\left<u|i\frac{\partial u}{\partial k_{xc}}\right>\right)=\frac{\partial\mathcal{E}}{\partial k_{xc}}+\dot{k}_{xc}\left(\frac{\partial A_y}{\partial k_{xc}}-\frac{\partial A_x}{\partial k_{yc}}\right).
\end{split}
\label{vyc}
\end{equation}
\end{widetext}
Combining Eqs.~\eqref{kc}, \eqref{vxc} and \eqref{vyc}, we get the more familiar expressions
\begin{align}
\dot{\mathbf{k}}_c&=eE
\\ 
\dot{\mathbf{r}}&=\frac{\partial \mathcal{E}}{\partial \mathbf{k}_c}+\dot{\mathbf{k}}_c\times\left(\nabla\times\mathbf{A}\right).
\end{align}

\subsection{Comparison to Ref.~\onlinecite{Cao20}}

We have employed a simple semiclassical description of the exciton that considers separate wavepackets for the electron and hole.
Reference~\onlinecite{Cao20} instead extended the single particle formalism for semiclassical dynamics to an exciton.  
In this case, the initial state is given by
\begin{equation}
\ket{\Psi(t=0)}=\int d\mathbf{K}\,a(\mathbf{K})\ket{\Phi_0(\mathbf{K})}
\label{exciton0}
\end{equation}
where $\mathbf{K}$ is the COM momentum, $a(\mathbf{K})=|a(\mathbf{K})|e^{-i\gamma(\mathbf{K},t)}$ and $|a(\mathbf{K})|$ centered at $\mathbf{K}=\mathbf{K}_c$.
The exciton ground state $\left|\Phi_0(\mathbf{K})\right>$ for a given COM momentum is
\begin{equation}
\ket{\Phi_0(\mathbf{K})}=\sum_{\mathbf{k}}C_{\mathbf{k}}(\mathbf{K})\left|\phi^{e,\uparrow}_{\mathbf{K},\mathbf{k}}\right>\otimes\left|\phi^{h,\uparrow}_{\mathbf{K},-\mathbf{k}}\right>.
\end{equation}
 At a later time $t$, this system is described by
\begin{equation}
\ket{\Psi(t)}=\int d\mathbf{K}a(\mathbf{K},t)\ket{\Phi(\mathbf{K},t)}
\end{equation}
where $\ket{\Phi(\mathbf{K},t)}=\sum_{\mathbf{k}}C_{\mathbf{k}}(\mathbf{K},t)\left|\phi^{e,\uparrow}_{\mathbf{K},\mathbf{k}}\right>\otimes\left|\phi^{h,\uparrow}_{\mathbf{K},-\mathbf{k}}\right>$. 
In order to study the dynamics of this system, we can again employ time-dependent variational principle with the Lagrangian of Eq.~\eqref{eq:Lagrangian} for  $H=H_0+\mathbf{E}\cdot\left(\mathbf{r}_e-\mathbf{r}_h\right)$ and $H_0=H^e\otimes\mathds{1}^h+\mathds{1}^e\otimes H^h+V(\mathbf{r}_e-\mathbf{r}_h)$.
Here, we can calculate the expectation value $\left<\Psi|\mathbf{r}_e-\mathbf{r}_h|\Psi\right>$  using
\begin{equation}
\begin{split}
\left<{\mathbf{r}_e,\mathbf{r}_h}|{\Phi(\mathbf{K},t)}\right>=e^{-i\frac{\mathbf{K}}{2}\cdot\left(\mathbf{r}_e+\mathbf{r}_h\right)}&\sum_{\mathbf{k}}C_{\mathbf{k}}(\mathbf{K},t)e^{-i\mathbf{k}\cdot\left(\mathbf{r}_e-\mathbf{r}_h\right)}
\\&\times \left|u^{e,\uparrow}_{\mathbf{K},\mathbf{k}}\right>\otimes\left|u^{h,\uparrow}_{\mathbf{K},-\mathbf{k}}\right>
\end{split}
\end{equation}
where $\left|u^{e/h,\uparrow}_{\mathbf{K},\mathbf{k}}\right>$ is cell-periodic part of Bloch wavefunctions of $H^{e/h}$ with momentum $\mathbf{k}_{e/h}={\frac{\mathbf{K}}{2}}+\mathbf{k}$. 
Similar to the technique employed in Ref.~\onlinecite{Cao20}, we can express 
\begin{equation}
\left<{\Phi(\mathbf{K},t)}|\mathbf{r}_e-\mathbf{r}_h|{\Phi(\mathbf{K},t)}\right>=\mathbf{A}^1(\mathbf{K},t)-\mathbf{A}^0(\mathbf{K},t)
\end{equation}
where $\mathbf{A}^\alpha(\mathbf{K},t)=i\left<{\Phi(\mathbf{K},t,\alpha)}|\nabla_{\mathbf{K}}|{\Phi(\mathbf{K},t,\alpha)}\right>$ and $\left|{\Phi(\mathbf{K},t,\alpha)}\right>=e^{-i(\alpha-\frac{1}{2})\mathbf{K}\cdot\left(\mathbf{r}_e-\mathbf{r}_h\right)}\left|{\Phi(\mathbf{K},t)}\right>$. 

If we assume the adiabaticity condition, $E$ does not change the exciton  eigenstate for a given $\mathbf{K}$.
As a result  $\left|{\Phi(\mathbf{K},t)}\right>=\left|{\Phi_0(\mathbf{K})}\right>$ and thus
\begin{equation}
\left<{\Psi(t)}|H_0+\mathbf{E}\cdot\left(\mathbf{r}_e-\mathbf{r}_h\right)|\Psi{(t)}\right>=E_0(\mathbf{K}_c)+\mathbf{E}\cdot\mathbf{D}(\mathbf{K}_c)
\end{equation}
where $\mathbf{D}((\mathbf{K}_c)=\mathbf{A}^1(\mathbf{K}_c)-\mathbf{A}^0(\mathbf{K}_c)$ is referred to as dipole curvature  and $\mathbf{K}_c=\mathbf{K}_c(t)=\int d\mathbf{K}|a(\mathbf{K},t)|^2\mathbf{K}$ is the mean of distribution at time $t$ and $E_0(\mathbf{K}_c)$ is the ground state energy of the exciton with COM momentum $\mathbf{K}_c$. 
Following the same steps as the single-particle case, it can be shown 
\begin{equation}
\begin{split}
\dot{\mathbf{R}}_c&=-\nabla_{\mathbf{K}_c}E_0(\mathbf{K}_c)-\nabla_{\mathbf{K}_c}\left(\mathbf{E}\cdot\mathbf{D}(\mathbf{K}_c)\right)\\&\quad\quad+\dot{\mathbf{K}}_c\times\left(\nabla_{\mathbf{K}_c}\times\mathbf{A}(\mathbf{K}_c)\right)\\
\dot{\mathbf{K}}_c&=0
\end{split}
\end{equation}
where $\mathbf{A}(\mathbf{K}_c)=i\left<{\Phi_0(\mathbf{K}_c)}|\nabla_{\mathbf{K}}|{\Phi_0(\mathbf{K}_c)}\right>$ is the Berry connection of exciton. 
Reference~\onlinecite{Cao20} found that the dipole curvature $\nabla_{\mathbf{K}_c}\left(\mathbf{E}\cdot\mathbf{D}(\mathbf{K}_c)\right)$ usually points in the  direction $\hat{z}\times\mathbf{K}_c$ for a simple 2D system with finite Berry curvature. 
Hence, in addition to the exciton Berry curvature, the dipole curvature term $\nabla_{\mathbf{K}_c}\left(\mathbf{E}\cdot\mathbf{D}(\mathbf{K}_c)\right)$  also gives rise to an anomalous transverse drift. 

\subsection{Comparison to exact dynamics simulation}

A key assumption of the above derivation is adiabaticity, so that for a given COM momentum $\mathbf{K}$, the exciton always remains in its ground state
\begin{equation}
\left|{\Phi(\mathbf{K},t)}\right>=\left|{\Phi_0(\mathbf{K})}\right>=\sum_{\mathbf{k}}C_{\mathbf{k}}(\mathbf{K})\left|\phi^{e,\uparrow}_{\mathbf{K},\mathbf{k}}\right>\otimes\left|\phi^{h,\uparrow}_{\mathbf{K},-\mathbf{k}}\right>.
\end{equation}
As a result, the expectation values of relative momentum $\mathbf{k}$ and relative position $\mathbf{r}_e-\mathbf{r}_h$ remain fixed during the evolution if $\dot{\mathbf{K}}_c=0$. 
This condition does not allow the difference in Berry curvature of electron and hole band to affect the COM motion and corresponds to the deep harmonic regime where the exciton is stuck at its equilibrium position both in $k$ and $r$ space.

In our exact dynamics, we start with a wavepacket in COM space similar to the one described in Eq.~\eqref{exciton0} and then we evolve it numerically. 
Hence, in our case we are not imposing this adiabaticity condition. 
Accordingly, the only way $k$ and $r$ can change in the absence of a net COM force is if the applied electric field mixes the ground state with other exciton states or other continuum states. 
The exciton remains bounded as long as as all states involved in the mixture are bounded.
Mixing with continuum states would dissociate the exciton before it can traverse the full Brillouin zone.

\section{1D Model}
\label{app:1D}

\begin{figure}[t]
    \centering
    \includegraphics[scale=0.5]{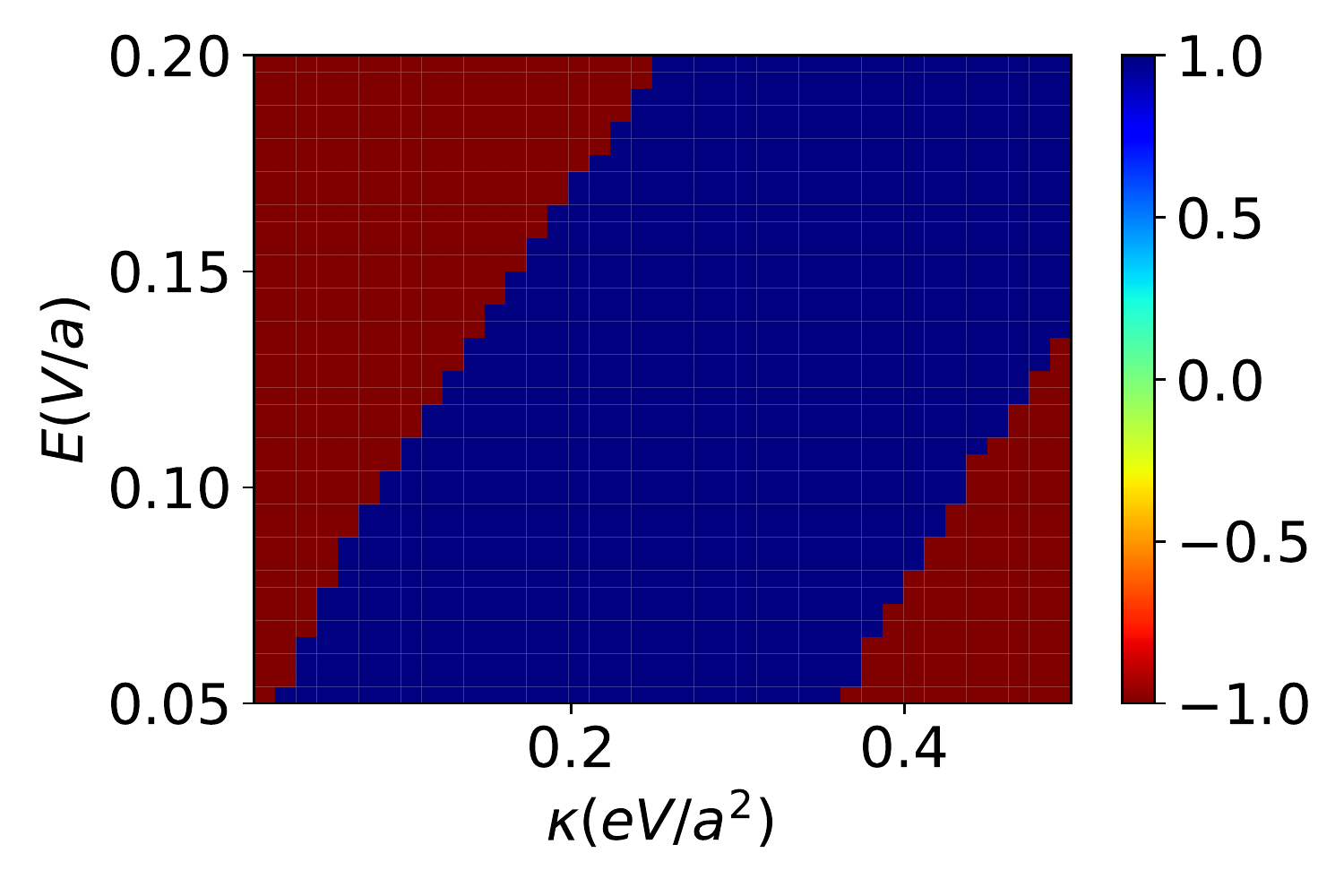}
    \caption{{\it Phase diagram for 1D model}: $\text{Sign}(x_\text{max}-x_\text{eq})$ for different values of $E$ and $\kappa$ as obtained from Eq.~\eqref{xmax} for $J_{1D}=0.08eV$. When $x_\text{max}-x_\text{eq}$ is negative, the particle cannot reach its equilibrium position.
    We see this occurs even at large $\kappa$, contrary to semiclassical predictions.}
    \label{fig:signxmax}
\end{figure}

\subsection{Perturbation theory in $J_\text{1D}/(\kappa a^2)$}

Consider the following one-band tight binding model with nearest neighbor hopping and a harmonic potential trap:
\begin{widetext}
\begin{align}
H_\text{1D} &= \sum_n \frac{J_\text{1D}}{2} \left( \ket{n}\bra{n+1}+ \ket{n}\bra{n-1}\right) \notag 
+\sum_n \left(\frac{1}{2} \kappa a^2\,\hat{n}^2 + eEa \,\hat{n} \right) \ket{n} \bra{n} 
\\ =& \left( \begin{array}{ccccc} 
2\kappa a^2- 2 eE a & \frac{J_\text{1D}}{2} & 0 & 0 &0 \\  \frac{J_\text{1D}}{2} & \frac{1}{2}\kappa a^2- eE a &  \frac{J_\text{1D}}{2}  & 0 & 0 \\ 0 &  \frac{J_\text{1D}}{2}  & 0 &  \frac{J_\text{1D}}{2}  & 0 \\ 0 & 0 &  \frac{J_\text{1D}}{2}  &\frac{1}{2} \kappa a^2+ eE a &  \frac{J_\text{1D}}{2}  \\ 0 & 0 & 0 &  \frac{J_\text{1D}}{2}  & 2\kappa a^2+ 2 eE a 
\end{array}\right).
\end{align}
In the second line, we truncate the Hamiltonian at states $\ket{n=\pm 2}$.
To second order in $J_\text{1D}/(\kappa a^2)$, the eigenvalues and eigenstates are 
\begin{align}
E_n &=\frac{1}{2} \kappa a^2 n^2 + eEa n  +\left( \frac{J_\text{1D}}{2} \right)^2 \left( \frac{1}{\kappa a^2 n + eEa-\frac{1}{2}\kappa a^2}-\frac{1}{\kappa a^2 n + eEa +\frac{1}{2} \kappa a^2} \right) 
\\ &= \frac{1}{2}\kappa a^2 n^2 + eEan +\left( \frac{J_\text{1D}}{2} \right)^2\frac{4\kappa a^2}{4(\kappa a^2 n + eEa)^2 -\kappa^2 a^4}
\end{align}
and

\begin{align}
\ket{\tilde{n}} 
= & \ket{n} \left( 1 - \frac{1}{2} \left( \frac{J_\text{1D}}{2} \right)^2 \sum_{\sigma=\pm1} \frac{1}{\left(\frac{1}{2} \kappa a^2+\sigma (n\kappa a^2+eEa)\right)^2} \right)-  
 \frac{J_\text{1D}}{2} \sum_{\sigma=\pm1} \ket{n+\sigma} \left( \frac{1}{\frac{1}{2}\kappa a^2+ \sigma (n \kappa a^2+ eEa) }\right) \\&+\notag 
\quad \frac{1}{2}  \left( \frac{J_\text{1D}}{2} \right)^2 \sum_{\sigma=\pm1} \ket{n+2\sigma} \left( \frac{1}{\frac{1}{2}\kappa a^2+ \sigma (n \kappa a^2+ eEa)  } \cdot \frac{1}{\kappa a^2+\sigma (n\kappa a^2 +eEa) } \right).
\end{align}

We initialize the system at $t=0$ in the ground state for $E=0,$
\begin{equation}
\ket{\psi(t=0)}=\left( 1 -  \frac{J_\text{1D}^2}{\kappa^2 a^4} \right)\ket{0} 
- \frac{J_\text{1D}}{\kappa a^2}  \ket{1}-\frac{J_\text{1D}}{\kappa a^2} \ket{-1}
 +  \frac{J_\text{1D}^2}{4\kappa^2 a^4}  \ket{2}+\frac{J_\text{1D}^2}{4\kappa^2 a^4} \ket{-2} .
\end{equation}
Rewriting the position eigenstates in terms of energy eigenstates $\ket{\tilde{n}}$, such that $H\ket{\tilde{n}}=E_n\ket{\tilde{n}}$, we have
\begin{align}
\ket{0} &= \left( 1-\left(\frac{J_\text{1D}}{2}\right)^2 \left[ \frac{2}{(\kappa a^2+ 2eEa)^2} + \frac{2}{(\kappa a^2- 2eEa)^2} \right] \right) \ket{\tilde{0}}  + \frac{J_\text{1D}}{\kappa a^2+2 eEa} \ket{\tilde{1}} +  \frac{J_\text{1D}}{\kappa a^2-2eEa} \ket{-\tilde{1}} 
\\ \ket{+1} &= \ket{\tilde{1}} - \frac{J_\text{1D}}{\kappa a^2+ 2eEa} \ket{\tilde{0}} +  \frac{J_\text{1D}}{3\kappa a^2+ 2eEa} \ket{\tilde{2}} +\mathcal{O}(J_\text{1D}^2)
\\ \ket{-1} &= \ket{-\tilde{1}} -\frac{J_\text{1D}}{\kappa a^2- 2eEa} \ket{\tilde{0}} +  \frac{J_\text{1D}}{3\kappa a^2+ 2eEa} \ket{-\tilde{2}} + \mathcal{O}(J_\text{1D}^2)
\\ \ket{+2} &= \ket{\tilde{2}} + \mathcal{O}(J_\text{1D})
\\ \ket{-2} &= \ket{-\tilde{2}} +\mathcal{O}(J_\text{1D}).
\end{align}
In addition to $J_\text{1D}\ll\kappa$, we have also assumed that $J_\text{1D}\ll|n\kappa\pm2E|$ for all $n$.
Ignoring $O(J_\text{1D}^2)$ terms, we find
\begin{equation}
\ket{\psi(t=0)}\approx\ket{\tilde{0}}-  \frac{2J_\text{1D}eEa}{\kappa a^2(\kappa a^2+2e Ea)}\ket{\tilde{1}} + \frac{2J_\text{1D}eEa}{\kappa a^2(\kappa a^2-2e Ea)} \ket{-\tilde{1}} 
\end{equation}
Evolving $\ket{\psi}$ according to $H$ with $E\neq 0$ we find
\begin{equation}
\begin{split}
\ket{\psi(t)}\approx e^{-i E_0 t} \left( \ket{\tilde{0}}- e^{-i\left(E_1-E_0\right)t}   \frac{2J_\text{1D}eEa}{\kappa a^2(\kappa a^2+2 eEa)} \ket{\tilde{1}} + e^{-i\left(E_{-1}-E_0\right)t}   \frac{2J_\text{1D}eEa}{\kappa a^2(\kappa a^2-2 eEa)} \ket{-\tilde{1}}\right)
\end{split}
\end{equation}
with position expectation value
\begin{equation}
\begin{split}
\bra{\psi(t)} a\hat{n} \ket{\psi(t)} =&a\bra{\tilde{0}}\hat{n} \ket{\tilde{0}} +a\left(\frac{2J_\text{1D}eEa}{\kappa a^2(\kappa a^2+2e Ea)}\right)^2\bra{\tilde{1}}\hat{n} \ket{\tilde{1}}+a\left(\frac{2J_\text{1D}eEa}{\kappa a^2(\kappa a^2-2e Ea)}\right)^2\bra{-\tilde{1}} \hat{n}  \ket{-\tilde{1}} \\&-2a\cos\left((E_1-E_0)t\right)\frac{2J_\text{1D}eEa}{\kappa a^2(\kappa a^2+2 eEa)}\bra{\tilde{0}} \hat{n}  \ket{\tilde{1}}+2a\cos\left((E_{-1}-E_0)t\right)\frac{2J_\text{1D}eEa}{\kappa a^2(\kappa a^2-2 eEa)}\bra{\tilde{0}} \hat{n}  \ket{-\tilde{1}}.
\end{split}
\end{equation}
Plugging in the position expectation values of the energy eigenstates
\begin{align}
    \bra{\psi(t)}a\hat{n}\ket{\psi(t)} &= \frac{J_\text{1D}^2}{\kappa^2 a^4} a \left( \frac{1 + 4n_\text{eq}^2 + 4 n_\text{eq} \cos \left( \left[n_\text{eq} + \frac{1}{2} \right] \kappa a^2 t\right)}{\left(1+2 n_\text{eq}\right)^2} - \frac{1+ 4n_\text{eq}^2 -4 n_\text{eq} \cos\left( \left[ n_\text{eq} - \frac{1}{2} \right] \kappa a^2 t\right)}{\left( 1- 2 n_\text{eq}\right)^2} \right)
\end{align}
where we have defined $n_\text{eq}=x_\text{eq}/a=eE/\kappa a$.
The above can be rewritten as
\begin{align}
	\bra{\psi(t)} a\hat{n} \ket{\psi(t)} =&4\frac{J_\text{1D}^2}{\kappa^2 a^4} x_\text{eq} \left( \frac{\cos\left(\left[ n_\text{eq} + \frac{1}{2} \right] \kappa a^2 t\right)-1}{(1+2n_\text{eq})^2} + \frac{\cos\left(\left[n_\text{eq}-\frac{1}{2} \right] \kappa a^2 t \right)-1 }{(1 - 2n_\text{eq})^2} \right)
\end{align}
\end{widetext}

The maximum amplitude $|\langle a\hat{n}\rangle| =x_\text{max}$ corresponds to both cosines taking value $-1$ (note it is not always possible to simultaneously maximize both cosines):
\begin{equation}\begin{split}
x_\text{max}
&=\frac{J_\text{1D}^2}{\kappa^2 a^4} a\frac{16n_\text{eq}}{(1+2n_\text{eq})^2(1-2n_\text{eq})^2}.
\end{split}
\label{xmax}
\end{equation}
When $x_\text{max}<x_\text{eq}$, the system never reaches its equilibrium value and always experiences a net force.
In the context of excitons, this implies there is a regime of large $\kappa$ for which the electron and hole never reach their equilibrium separation and therefore undergo Bloch oscillations.  
Fig.~\ref{fig:signxmax} plots $\text{Sign}(x_\text{max}-x_\text{eq})$ for $J_{1D}=0.08eV$.
The analysis in this appendix relies on perturbation theory; it does not apply to regions of the phase diagram for which $J_\text{1D}/\kappa a^2$ and $J_\text{1D}/|n\kappa a^2\pm 2E a|$ are not small.  

\begin{figure}[t]
    \centering
    \includegraphics[scale=0.27]{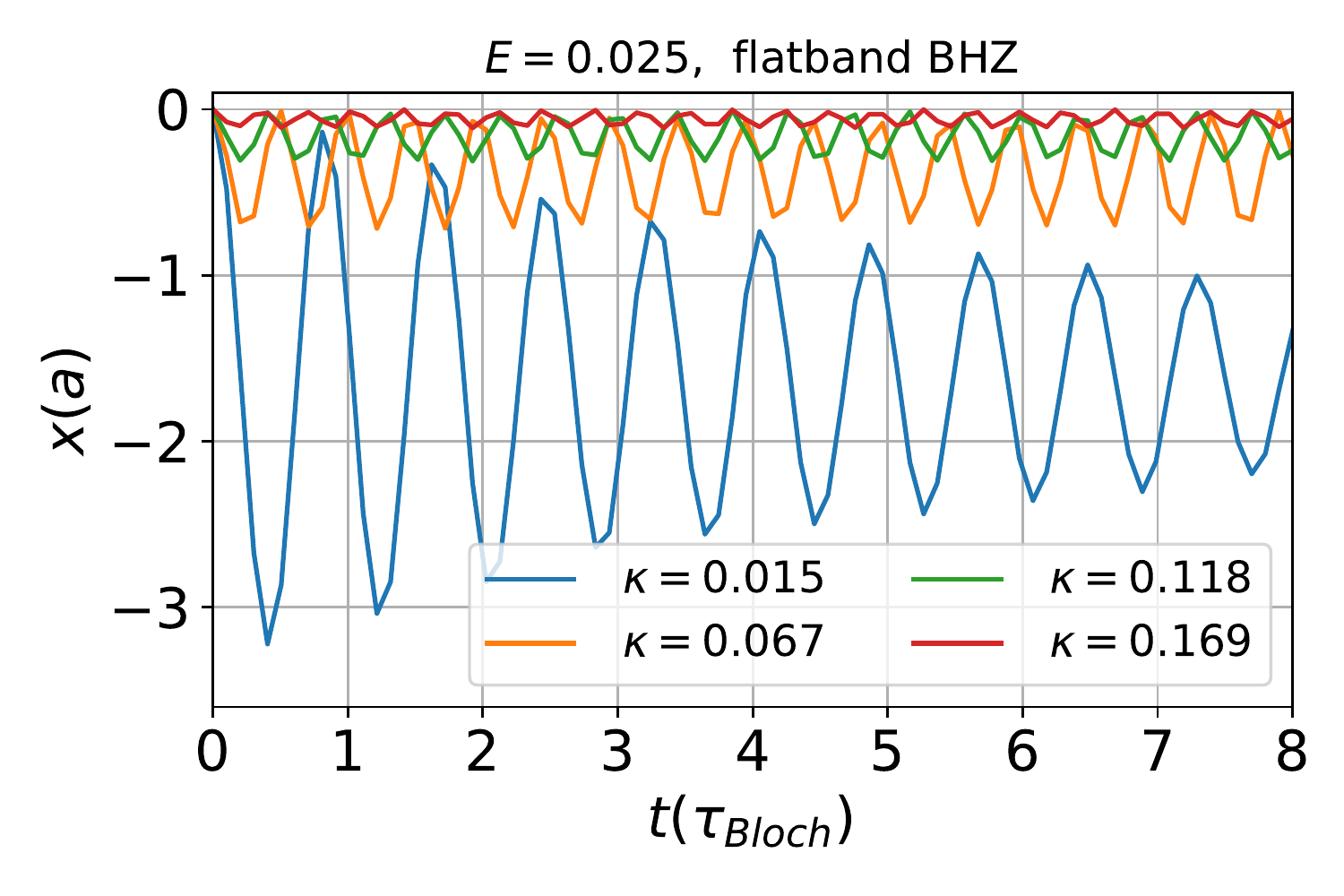}
    \includegraphics[scale=0.27]{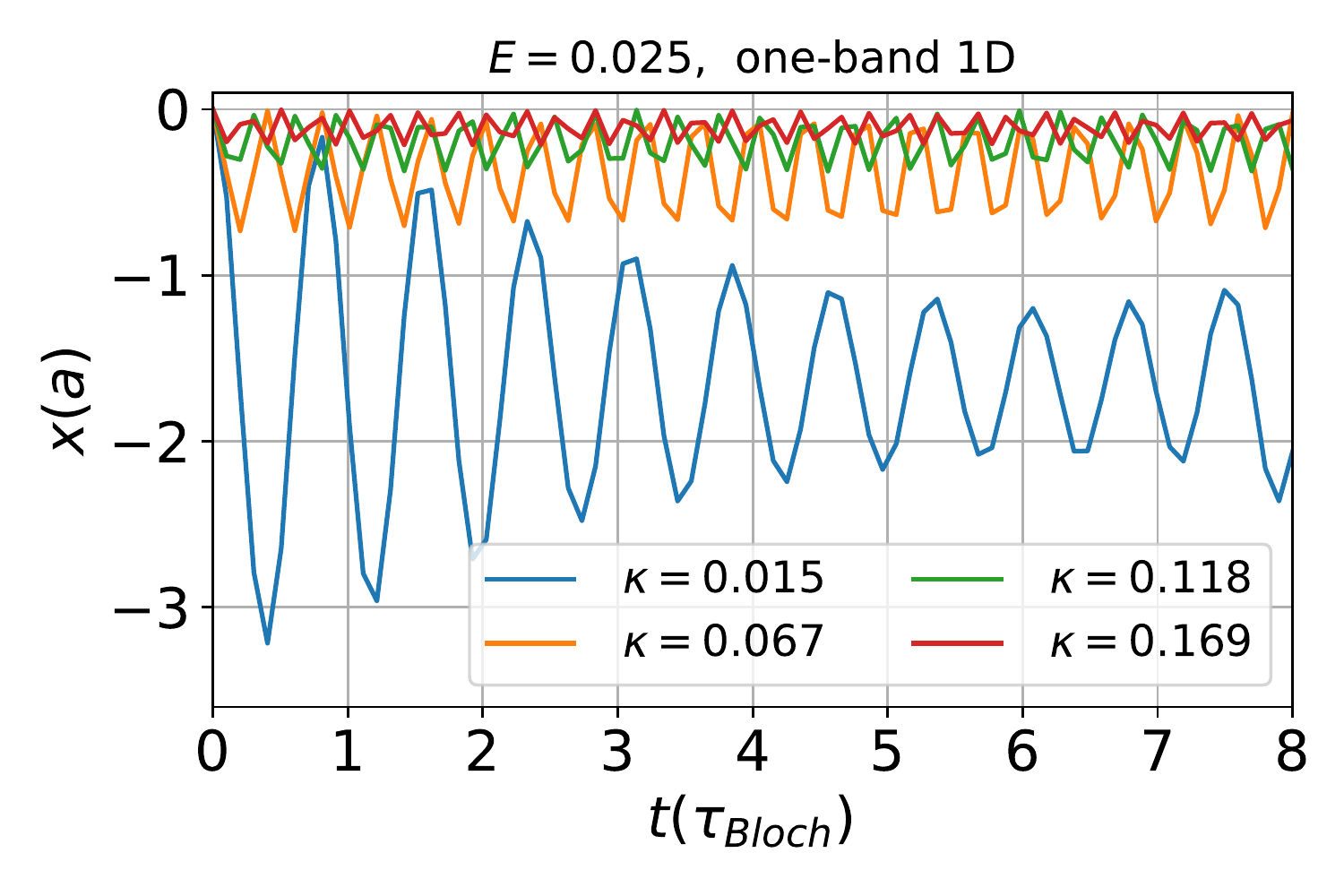}
    \includegraphics[scale=0.27]{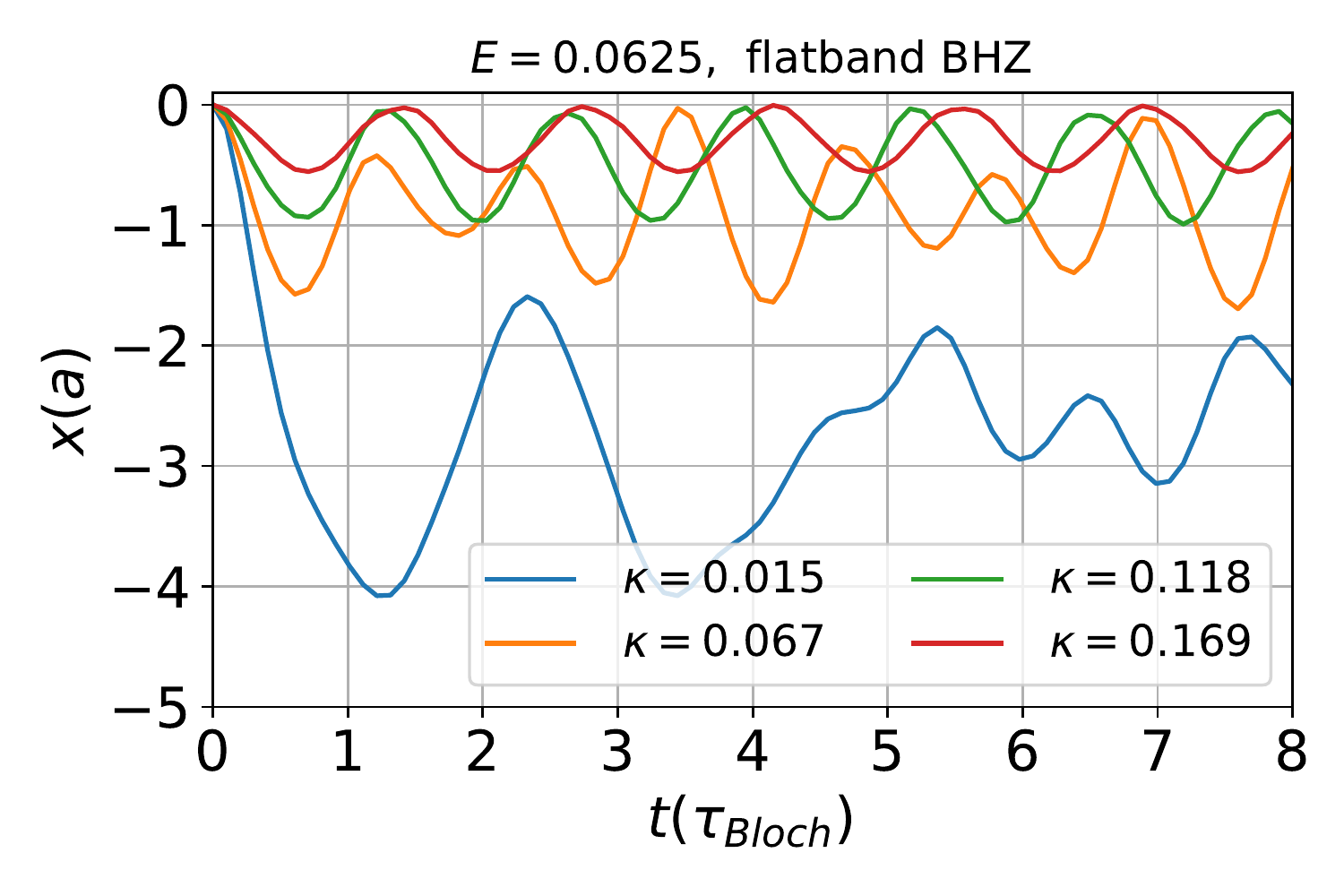}
     \includegraphics[scale=0.27]{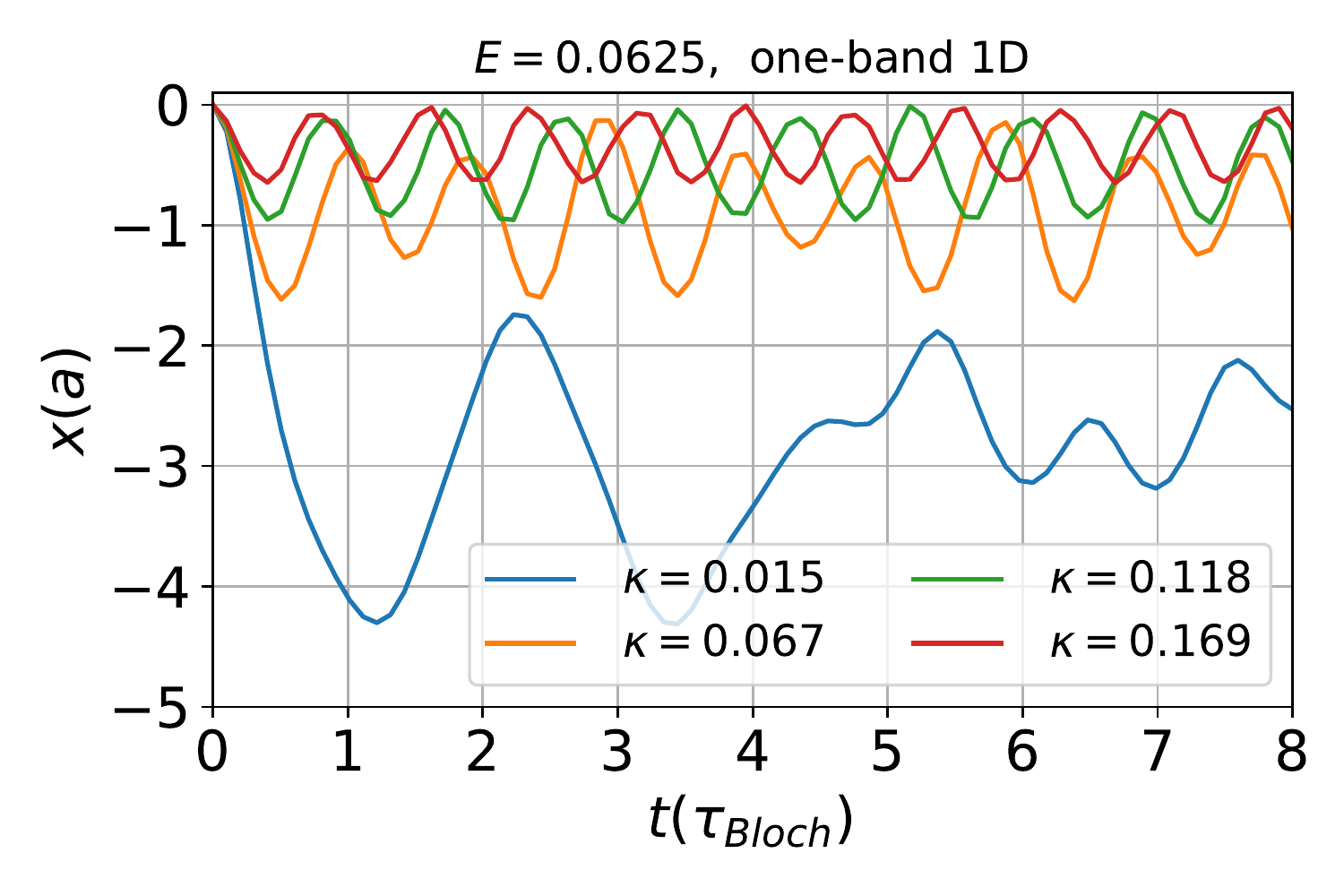}
    \includegraphics[scale=0.27]{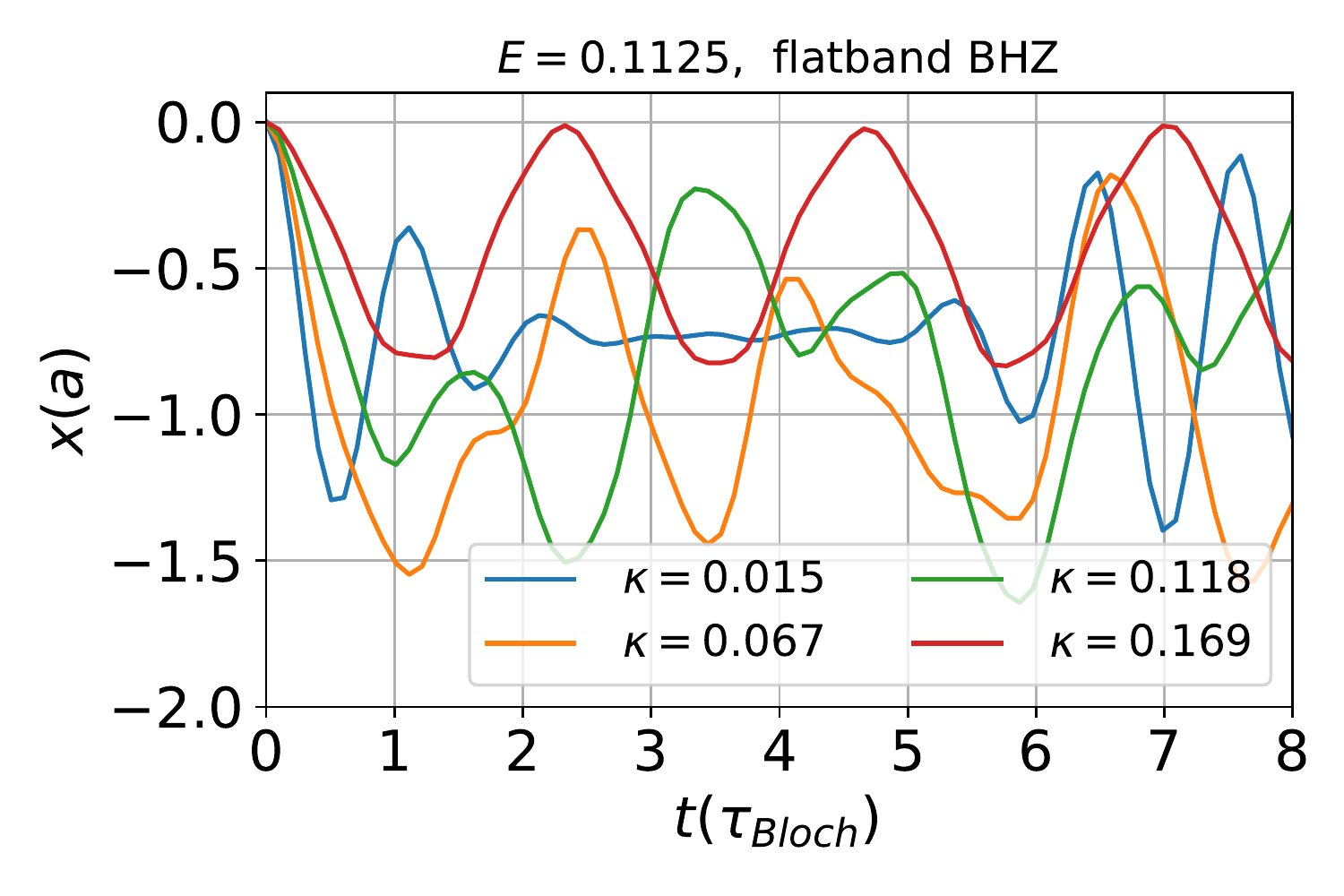}
    \includegraphics[scale=0.27]{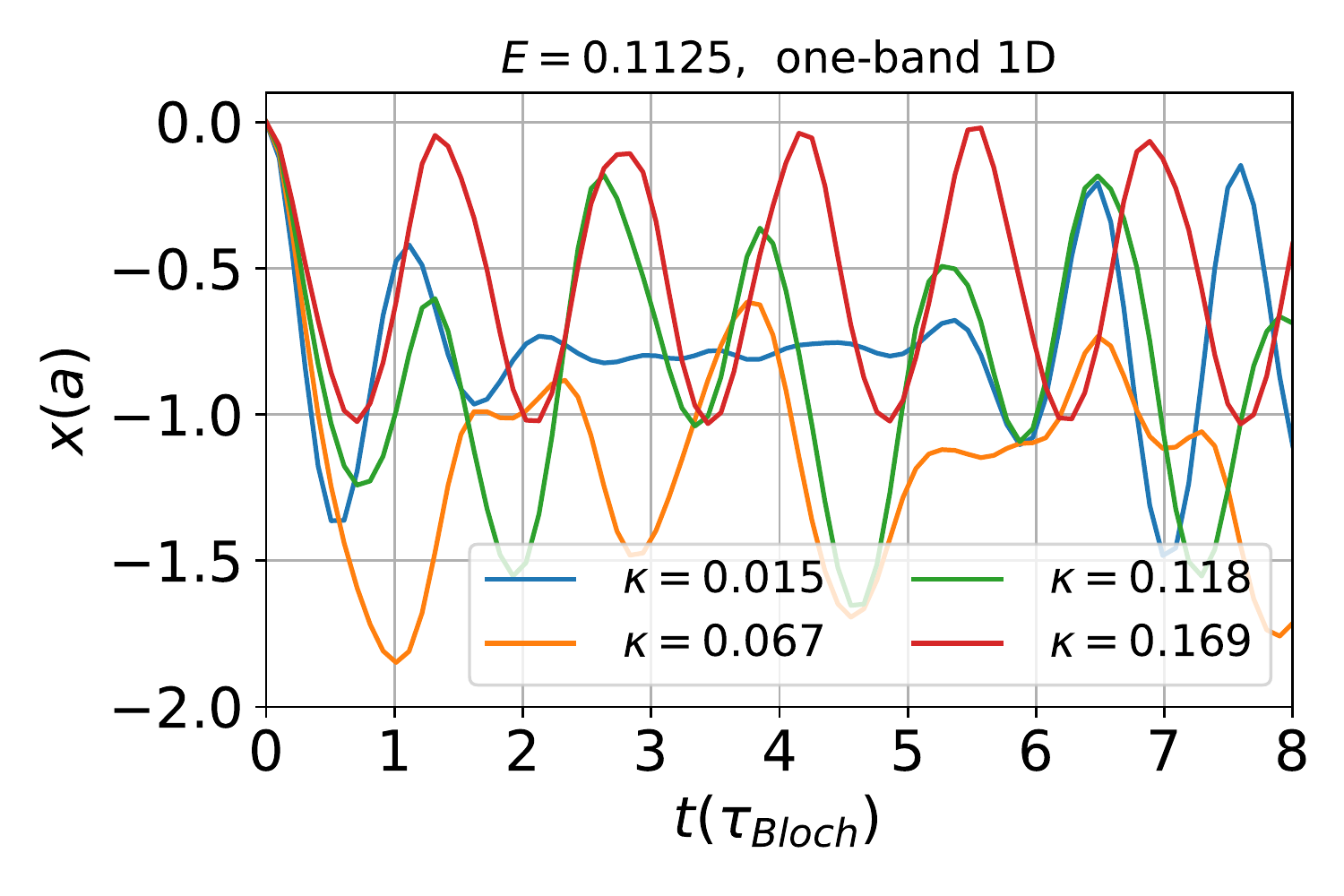}
    \caption{Relative position $x$ as a function of $t$ obtained from exact dynamics for the 1D version of Eq.~\eqref{HBHZcomfb} with COM momentum $K=0$ (left) and for $H_\text{1D}$ (right). }
    \label{fig:1Dmodel}
\end{figure}

\subsection{Comparison between one-band model and $H_\text{BHZ}^\text{FB}(k_y=0)$}

In order to verify that above 1D model can capture the dynamics of the exciton Hamiltonian used in our numerical simulations, we compare the evolution of relative coordinate $x$ for the two cases.
We compare dynamics according to $H_\text{1D}$ presented above and the 1D form of Eq.~\eqref{HBHZcomfb} with a harmonic potential and $\mathbf{K}=0$.
More specifically, the latter replaces each copy of $H_\text{BHZ}^\text{FB}(\mathbf{K},\mathbf{r})$ with the real-space version of $H_\text{BHZ}^\text{FB}(\mathbf{K}=0,k_y=0)$.  

Figure~\ref{fig:1Dmodel} shows qualitative agreement between the two cases. 
We note that large $\kappa$ suppresses the Bloch oscillations in both cases.

\section{Exact dynamics simulation details}\label{app:simulation}

In this appendix, we provide details of our numerical simulations.
We first explain our non-interacting Hamiltonian composed of two copies of the BHZ Hamiltonian~\cite{Bernevig06}, one for the electron and one for the hole, written in relative real space and COM momentum space.
We then explain our band flattening method.
Next, we describe our projection into the exciton Hilbert space and incorporating electron-hole interactions.
Finally, we describe the ground state preparation and its times evolution.

\subsection{Non-interacting Hamiltonian}

\begin{figure}[t]
\centering
\includegraphics[scale=0.8]{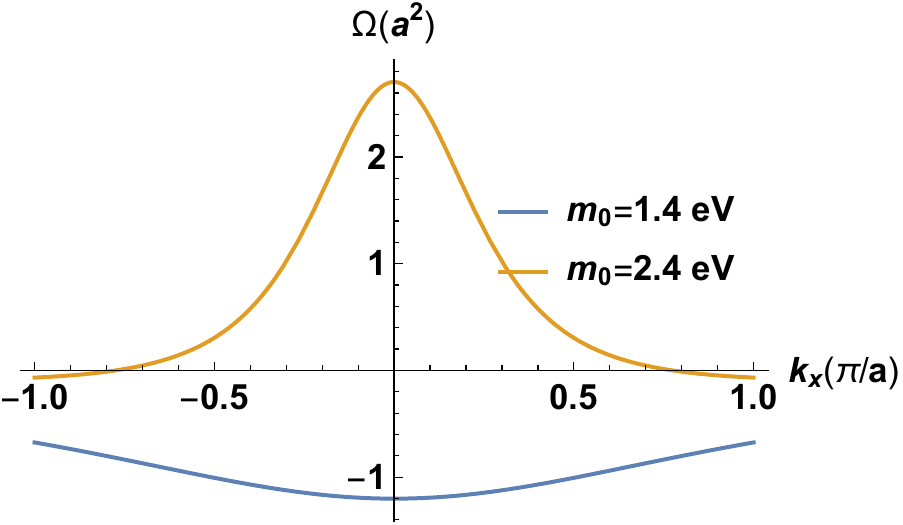}
\caption{{\it Berry curvature profile of $H_\text{BHZ}(k_y=0)$.} 
The Berry curvature profile of the upper band is plotted for the topological and trivial cases discussed in the main text, with $v_x=v_y=0.93$\,m/s, $b=1$\,eV.
}
\label{fig:BC-profile}
\end{figure}

	We consider two copies of the BHZ Hamiltonian
	\begin{align}
		H^\alpha_\text{BHZ}(\mathbf{k}_\alpha)&=\left(m_0-b\cos (k_x^\alpha a) -b\cos (k_y^\alpha a)\right)\sigma_z \notag 
		\\&\quad +v_x^\alpha \sin( k_x^\alpha a) \sigma_x+v_y^\alpha \sin (k_y^\alpha  a) \sigma_y
	\end{align}
	for $\alpha=e/h$ for the electron and hole, respectively. 
	In Fig.~\ref{fig:BC-profile} we plot the Berry curvature profile for the two values of $m_0$ used in the figures in the main text; $m_0=1.4$\,eV (blue curve) corresponds to topological bands with Chern number $\mathcal{C}=-1$, while $m_0=2.4$\,eV (yellow curve) corresponds to trivial bands with Chern number $\mathcal{C}=0$.  
	We introduce COM and relative coordinates:
	\begin{align}
		\mathbf{K}&=\mathbf{k}_h+\mathbf{k}_e, & \mathbf{k}&=(\mathbf{k}_e-\mathbf{k}_h)/2
	\end{align}
	in terms of which the electron and hole momenta can be written 
	\begin{align}
    \mathbf{k}_{e/h}&=\frac{\mathbf{K}}{2}\pm \mathbf{k}.
	\end{align}
	In our simulations, $\mathbf{K}$ is conserved and thus different $\mathbf{K}$ sectors are completely decoupled. 
	For a given $\mathbf{K}$, 
	\begin{widetext}
	\begin{align}
		&H^\alpha_{\text{BHZ}}\left(\mathbf{K} ,\mathbf{k}\right)=\left(m_0-\sum_{i}b\cos \left([\frac{K_i}{2}+k_i]a\right)\right)\sigma_z
		+\sum_{i}v^\alpha_i\sin \left([\frac{K_i}{2}+k_i]a\right)\sigma_i  
		\\ =& \left[m_0  -\sum_{i}b\cos \left(\frac{K_i a}{2}\right)\cos( k_i a) \mp\sin \left(\frac{K_i a}{2}\right)\sin (k_i a)\right]\sigma_z  
	 +\sum_{i}v^\alpha_i\left[\sin \left(\frac{K_i a}{2}\right)\cos(k_i a) \pm \cos \left(\frac{K_i a}{2}\right)\sin (k_i a)\right]\sigma_i.&
	\end{align}
	\end{widetext}

	We do a partial Fourier transform on $\mathbf{k}$ to write $H^\alpha_{\text{BHZ},\mathbf{K}}$ which is a tight-binding hamiltonian in relative position basis with nearest neighbor hopping only. Now, the Hilbert space is $\mathcal{H}_e\otimes\mathcal{H}_h\otimes\mathcal{H}_{\mathbf{r}}$ where $\mathcal{H}_{e/h}$ is the two-dimensional Hilbert space associated with electron/hole degrees of freedom and $\mathcal{H}_{\mathbf{r}}$ is the $N_x\times N_y$-dimensional Hilbert space spanned by the relative position eigenstates $\left|\mathbf{r}\right>$. 
	We can express any state in the full Hilbert space as 
	\begin{equation}
		\left|\psi\right>=\sum_{i,j=1,2}\sum_{\mathbf{r}}\alpha_{i,j}(\mathbf{r})\left|{e^i}\right>\otimes\left|{h^j}\right>\otimes\left|\mathbf{r}\right>
	\end{equation}
	where $\sum_{i,j=1,2}\sum_{\mathbf{r}}|\alpha_{i,j}(\mathbf{r})|^2=1$.
	Here, $\mathbf{r}=m\mathbf{a}_1+n\mathbf{a}_2$ where $\mathbf{a}_{i}$  are the lattice vectors of the underlying lattice.
	The full non-interacting tight-binding Hamiltonian at a fixed $\mathbf{K}$ is given by
	\begin{align}
		H_{\mathbf{K}}^\text{(0)}&=H^{e}_{\text{BHZ},\mathbf{K} }\otimes\mathds{1}_h +\mathds{1}_e\otimes H^{h}_{\text{BHZ},\mathbf{K}}.
		\label{HBHZ}
	\end{align}

	\subsection{Band flattening method}
	
	We now detail our band flattening procedure.  
	We modify each band so that the eigenstates (and Berry curvatures) are unchanged, but the bandwidth is significantly reduced.
	We then add back in a finite, cosine dispersion.
	This scheme only applies to a gapped Hamiltonian.
	
\begin{figure}
		  \centering
		   \includegraphics[scale=0.5]{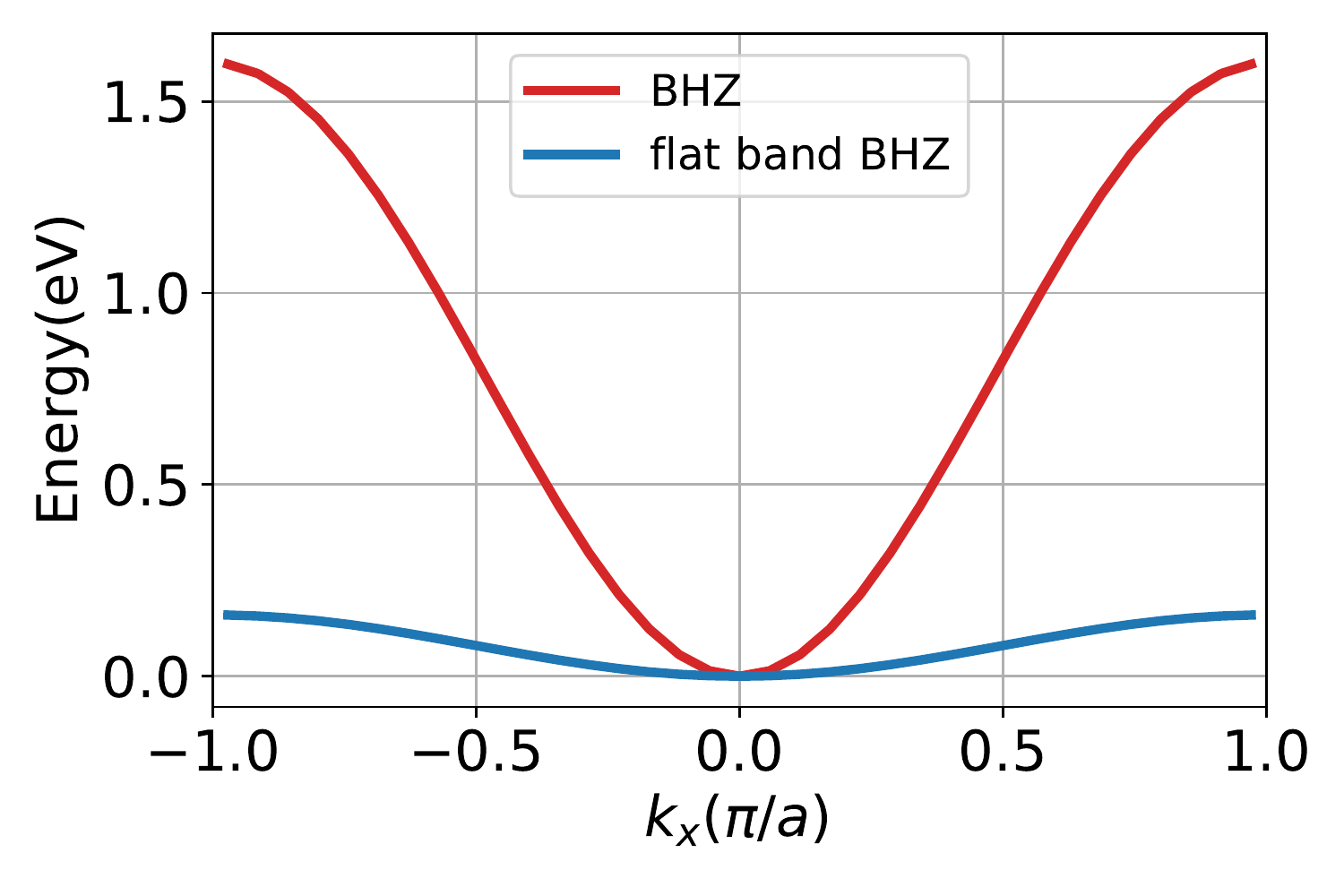}
		   \caption{Band dispersion of the single-particle $H_\text{BHZ}$ before and after the band flattening process. This flat band BHZ Hamiltonian has a bandwidth determined by the extra nearest-neighbor hopping added to $\tilde{H}_\text{BHZ}$.}
		   \label{fig:bandflat}
	\end{figure}

	We first flatten each single particle band completely by replacing all positive eigenvalues $\varepsilon_{\alpha}>0$ by the same number $E_0$, and all negative eigenvalues by the opposite constant $ -E_0.$
	In the eigenstate basis, ${H_\text{BHZ}\ket{\phi_\alpha}=\varepsilon_\alpha\ket{\phi_\alpha}}$, the transformation takes 
	\begin{align}
	    H_\text{BHZ}=\sum_{\alpha;\,\varepsilon_{\alpha}>0} \varepsilon_{\alpha} \ket{\phi_\alpha}\bra{\phi_\alpha} - \sum_{\alpha;\,\varepsilon_{\alpha}<0}\varepsilon_\alpha  \ket{\phi_\alpha}\bra{\phi_\alpha}
	\end{align}
	to the completely flattened
	\begin{align}
	    \tilde{H}_\text{BHZ}=E_0 \sum_{\alpha;\,\varepsilon_{\alpha}>0}  \ket{\phi_\alpha}\bra{\phi_\alpha} - E_0 \sum_{\alpha,\,\varepsilon_{\alpha}<0}  \ket{\phi_\alpha}\bra{\phi_\alpha}.
	\end{align}
    We can then perform a basis change to write $\tilde{H}_\text{BHZ}$ in relative position space.
    More explicitly, the algorithm implements the following steps:
	\begin{enumerate}
		\item Express $E_k=a_0+\sum_{m,n \neq 0}a_{mn  }\cos([mk_x+nk_y]a)$.
		\item Extract  $a_{mn}$ and introduce them as hopping between $\mathbf{r}_{mn}$ neighbors. The resulting Hamiltonian should give completely flat bands but with the same spinor structure at each $k$ point. 
	\end{enumerate}
	This scheme provides completely flat bands at the cost of non-local hopping. 
	In order to get a finite bandwidth, we scale $\tilde{H}_\mathbf{BHZ}$ by a matrix containing only nearest-neighbor hopping so that the bandwidth is directly proportional to these nearest neighbor terms, as shown in Fig.~\ref{fig:bandflat}. 
	
\begin{figure}
		\includegraphics[scale=0.5]{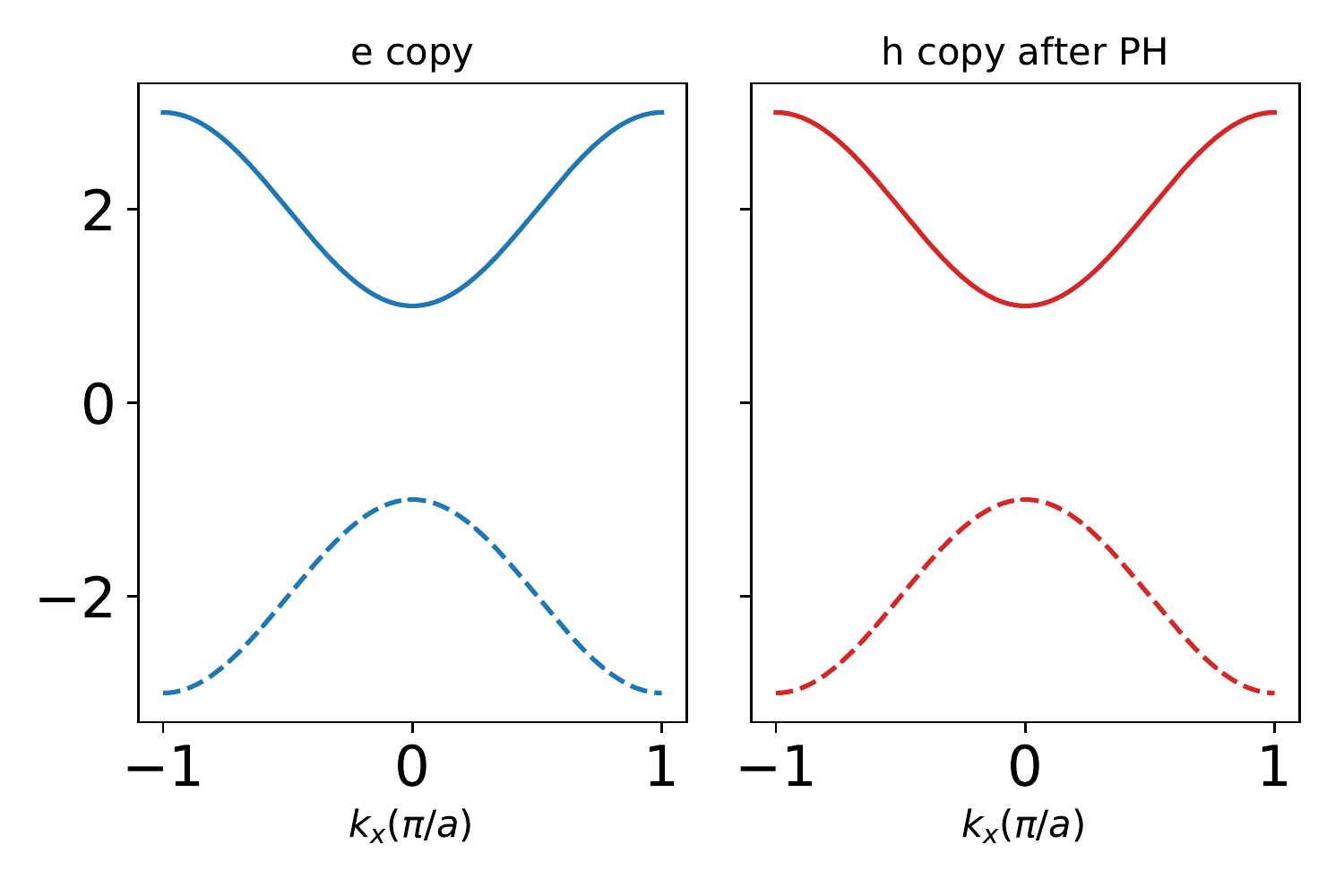}
		\includegraphics[scale=0.5]{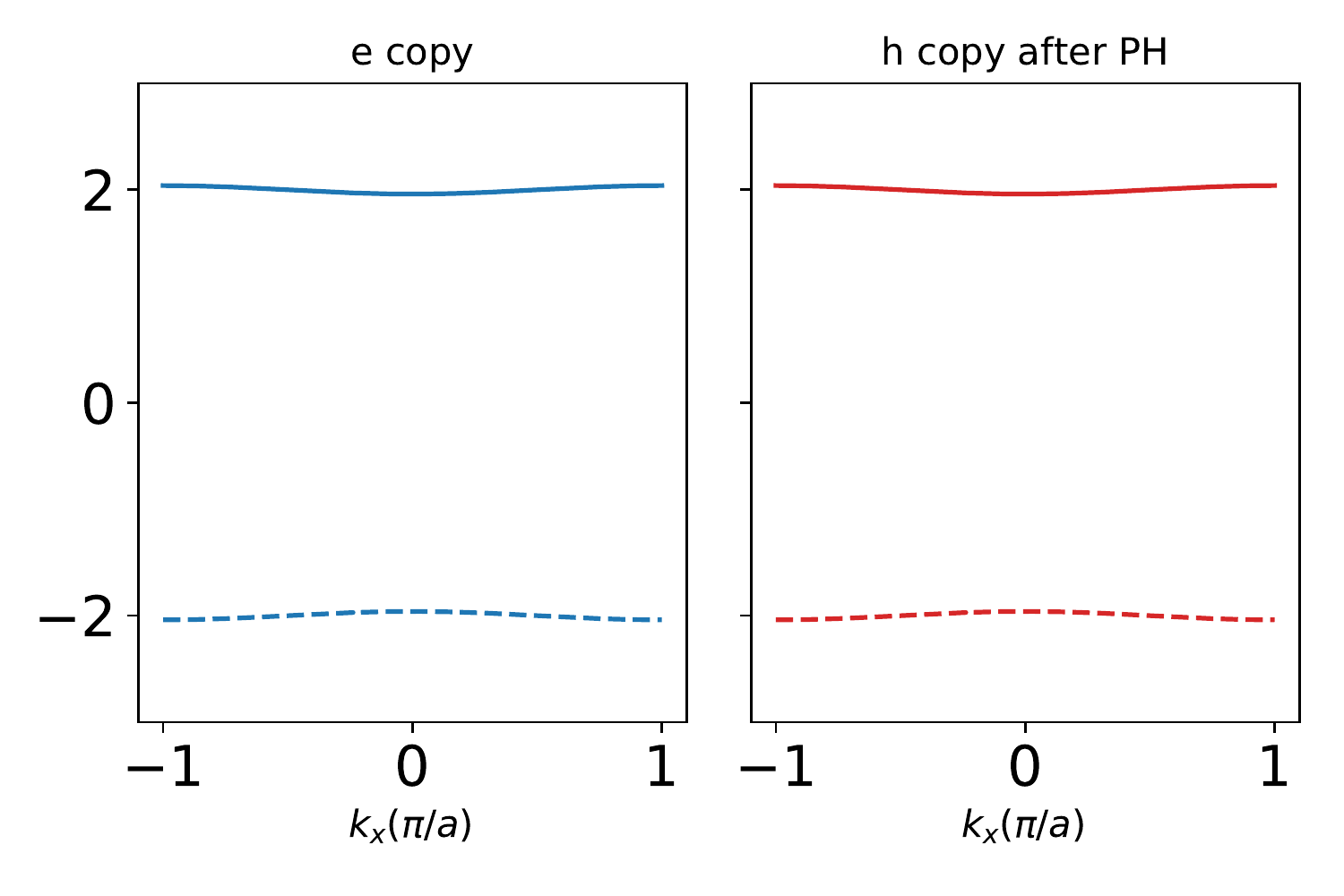}
		\includegraphics[scale=0.5]{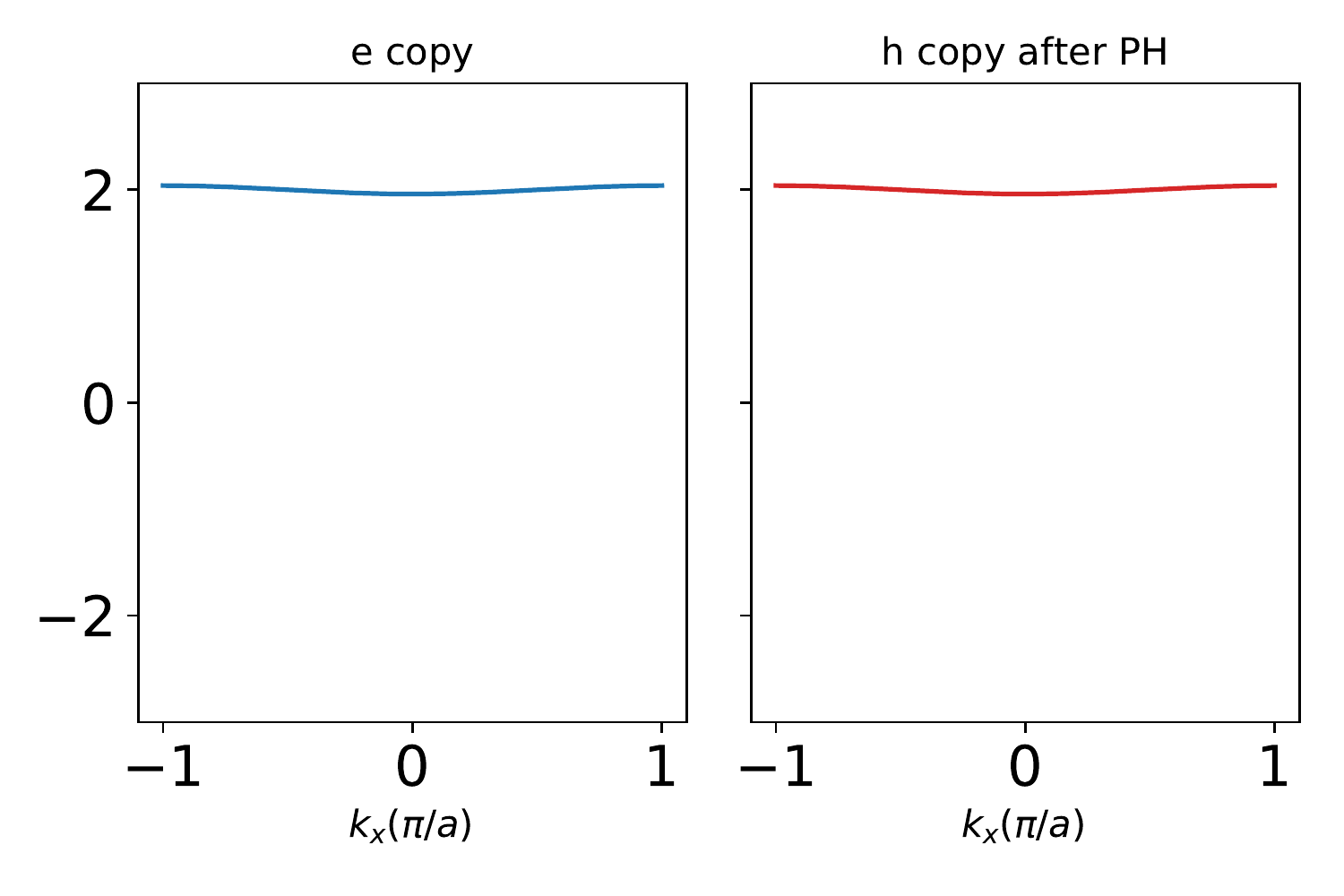}
		\caption{{\it Schematic of simulation procedure.} {\it Top}: We start with two copies of a two-band Hamiltonian. Bands shown in dashed line are fully occupied. An exciton is supposed to form between upper bands of electron and hole.  {\it Center}: Same bands after band flattening procedure. {\it Bottom}: We project into the exciton Hilbert space.  
		}
	\label{flowchart}
	\end{figure}
	
\subsection{Projection and interactions}

After band flattening, our new Hamiltonian is given by:
\begin{equation}
\tilde{H}^{(0)}_{\mathbf{K}}=\tilde{H}^e_{\text{BHZ},\mathbf{K}}\otimes\mathds{1}_h+\mathds{1}_e\otimes\tilde{H}^h_{\text{BHZ},\mathbf{K}}.
\label{hbhzbf}
\end{equation}
We want to project to the intervalley exciton Hilbert space, formed by both the electron and hole occupying the upper band of $H_\text{BHZ}.$
This is accomplished using the projectors $\hat{P}_{e/h}$
\begin{equation}
	\hat{P}_{e/h}=\sum_{\alpha,\mathcal{E}_\alpha>0} \left|\phi_{e/h}^\alpha\right>\left<\phi_{e/h}^\alpha\right|
\end{equation}
where $\ket{\phi_{e/h}^\alpha}$ are eigenstates of $H^{e/h}_\text{BHZ}.$
After projection, both valence and conduction bands (equivalently, hole and electron bands in our model) will have the same Berry curvature.
Figure~\ref{flowchart} illustrates the projection procedure.

We incorporate interactions using 
\begin{equation}
H_\text{int}= \hat{P}_{e}	\hat{P}_{h}\left(\sum_{\mathbf{r}}V(\mathbf{r})\mathds{1}_e\otimes\mathds{1}_h\otimes\left|\mathbf{r}\right>\left<\mathbf{r}\right|\right)\hat{P}_{h}\hat{P}_{e} .
\label{eq:intP}
\end{equation}
Note that including the projectors ensures that we only consider interactions between electrons and holes within the exciton Hilbert space.

\subsection{Ground state preparation and time-evolution}

For each $K$, our full Hamiltonian in absence of $E$ is given by:
\begin{equation}
    H^\text{ex}_{\mathbf{K}}=	\hat{P}_{e}\,	\hat{P}_{h}\,\tilde{H}^{(0)}_{\mathbf{K}}\,	\hat{P}_{h}\,\hat{P}_{e} +H_\text{int}
\end{equation}
where $\tilde{H}^{(0)}_{\mathbf{K}}(\mathbf{r})$ and $H_\text{int}(\mathbf{r})$ are defined in Eqs.~\eqref{hbhzbf} and \eqref{eq:intP}. 
Denote the ground state of this Hamiltonian by $\ket{\Phi_0(\mathbf{K})}$.
Beginning from $\ket{\Phi_0(\mathbf{K})}$, we evolve according to $H_\text{exciton}+H_E$ where
\begin{equation}
    H_E=\hat{P}_{e}\,\hat{P}_h\, \left(\sum_{\mathbf{r}} e\mathbf{E}\cdot\mathbf{r}\right)\, \hat{P}_{h}\,\hat{P}_{e}	.
\end{equation}
We repeat the same process for each $K$ on a grid of $27\times81$ points. At an arbitrary time, the full state of the system is given by:
\begin{equation}
    \ket{\psi(t)}=\sum_{\mathbf{K}}w(\mathbf{K})\ket{\Phi_t(\mathbf{K})}
\end{equation}
where ${w(\mathbf{K})=e^{-K^2\sigma_k^2}}$, ${\ket{\Phi_t(\mathbf{K})}=e^{-i\left(H^\text{ex}_{\mathbf{K}}+H_E\right)t}\ket{\Phi_0(\mathbf{K})}}$.
We choose a narrow wavepacket with $\sigma_K=\pi/15$ and a numerically smooth gauge such that the initial wavepacket is a coherent wavepacket in both $R$ and $K$ space. 
We then extract the COM position by performing a Fourier transform in $\text{COM}$ space. 
The observed $Y$ does not change qualitatively as we vary the wavepacket width $\sigma_K$ from $\pi/10$ to $\pi/20$.

\bibliography{exciton_ref.bib}

\end{document}